\documentclass[twocolumn]{aastex63}
\usepackage{graphicx,times}    
\usepackage{amssymb,amsmath}
\usepackage{amsfonts}
\usepackage{subfigure}
\usepackage{natbib}

\newcommand{\teff}{\ensuremath{T_{\mathrm {eff}}\,}}
\newcommand{\logg}{\ensuremath{{\mathrm {log}\, } g\,}}
\newcommand{\feh}{\ensuremath{[{\mathrm {Fe/H}}]\,}}

\shorttitle{Stellar abundances from DESI}
\shortauthors{Zhang et al.}

\begin{document}
\title{Determining Stellar Elemental Abundances from DESI Spectra with the Data-Driven Payne}
\author{Meng Zhang}
\affiliation{National Astronomical Observatories, Chinese Academy of Sciences, Beijing 100101, China\\}
\author{Maosheng Xiang}
\affiliation{National Astronomical Observatories, Chinese Academy of Sciences, Beijing 100101, China\\}
\affiliation{Institute for Frontiers in Astronomy and Astrophysics, Beijing Normal University, Beijing 102206, China\\}
\correspondingauthor{Maosheng Xiang}
\email{msxiang@nao.cas.cn}
\author{Yuan-Sen Ting}
\affiliation{Research School of Astronomy \& Astrophysics, Australian National University, Cotter Rd., Weston, ACT 2611, Australia\\}
\affiliation{School of Computing, Australian National University, Acton, ACT 2601, Australia\\}
\affiliation{Department of Astronomy, The Ohio State University, Columbus, OH 45701, USA.}
\affiliation{Center for Cosmology and AstroParticle Physics (CCAPP), The Ohio State University, Columbus, OH 43210, USA.\\}
\author{Jiahui Wang}
\affiliation{School of Astronomy and Space Science, University of Chinese Academy of Sciences, Beijing 100049, China\\}
\affiliation{National Astronomical Observatories, Chinese Academy of Sciences, Beijing 100101, China\\}
\author{Haining Li}
\affiliation{National Astronomical Observatories, Chinese Academy of Sciences, Beijing 100101, China\\}
\author{Hu Zou}
\affiliation{National Astronomical Observatories, Chinese Academy of Sciences, Beijing 100101, China\\}
\affiliation{School of Astronomy and Space Science, University of Chinese Academy of Sciences, Beijing 100049, China\\}
\author{Jundan Nie}
\affiliation{National Astronomical Observatories, Chinese Academy of Sciences, Beijing 100101, China\\}
\author{Lanya Mou}
\affiliation{National Astronomical Observatories, Chinese Academy of Sciences, Beijing 100101, China\\}
\affiliation{School of Astronomy and Space Science, University of Chinese Academy of Sciences, Beijing 100049, China\\}
\author{Tianmin Wu}
\affiliation{National Astronomical Observatories, Chinese Academy of Sciences, Beijing 100101, China\\}
\affiliation{School of Astronomy and Space Science, University of Chinese Academy of Sciences, Beijing 100049, China\\}
\author{Yaqian Wu}
\affiliation{National Astronomical Observatories, Chinese Academy of Sciences, Beijing 100101, China\\}
\author{Jifeng Liu}
\affiliation{National Astronomical Observatories, Chinese Academy of Sciences, Beijing 100101, China\\}
\affiliation{School of Astronomy and Space Science, University of Chinese Academy of Sciences, Beijing 100049, China\\}
\affiliation{Institute for Frontiers in Astronomy and Astrophysics, Beijing Normal University, Beijing 102206, China\\}
\affiliation{ WHU-NAOC Joint Center for Astronomy, Wuhan University, Wuhan, Hubei 430072, China\\}
\begin{abstract}
Stellar abundances for a large number of stars are key information for the study of Galactic formation history. Large spectroscopic surveys such as DESI and LAMOST take median-to-low resolution ($R\lesssim5000$) spectra in the full optical wavelength range for millions of stars. However, line blending effect in these spectra causes great challenges for the elemental abundances determination. Here we employ the {\sc DD-Payne}, a data-driven method regularised by differential spectra from stellar physical models, to the DESI EDR spectra for stellar abundance determination. Our implementation delivers 15 labels, including effective temperature $T_{\rm eff}$, surface gravity $\log~g$, microturbulence velocity $v_{\rm mic}$, and abundances for 12 individual elements, namely C, N, O, Mg, Al, Si, Ca, Ti, Cr, Mn, Fe, Ni. Given a spectral signal-to-noise ratio of 100 per pixel, internal precision of the label estimates are about 20~K for $T_{\rm eff}$, 0.05~dex for $\log~g$, and 0.05\,dex for most elemental abundances. These results are agree with theoretical limits from the Cr\'amer-Rao bound calculation within a factor of two. The Gaia-Enceladus-Sausage that contributes the majority of the accreted halo stars are discernible from the disk and in-situ halo populations in the resultant [Mg/Fe]-[Fe/H] and [Al/Fe]-[Fe/H] abundance spaces. We also provide distance and orbital parameters for the sample stars, which spread a distance out to $\sim$100~kpc. The DESI sample has a significant higher fraction of distant (or metal-poor) stars than other existed spectroscopic surveys, making it a powerful data set to study the Galactic outskirts. The catalog is publicly available.
\end{abstract}

\keywords{survey}
\section{Introduction}
\label{sec:intro}
Stellar abundances are fossil records of a star's birth environment. They thus serve as a powerful tool, in parallel with stellar orbits, to disentangle in-situ stellar populations and remnants of galaxy accretion events \citep[e.g.][]{Freeman2002, Frebel2005, Hawkins2015, BlandHawthorn2016, Helmi2020}. To deliver the abundances for a large sample of stars is one of the core goals of extensive spectroscopic surveys that aim to unraveling the formation and evolution history of our Galaxy \citep[e.g.][]{Gilmore2012, deJong2019, Takada2014, DeSilva2015, LiuC2020, Majewski2017,Kollmeier2017, Conroy2019}. 

Stellar abundances for individual elements are used to be derived from high-resolution ($R\gtrsim20,000$) spectra with a line-by-line analysis technique \citep[e.g.][]{Nissen2015, Jofre2019}. However, taking high-resolution spectra is expensive, and restricted to relatively small number of bright stars. Even the existed largest high-resolution spectroscopic surveys can only sample $O(10^5)$ stars \citep{Buder2022,Abdurrouf2022, Randich2022}, which constitutes only a fraction of $10^{-6}$ of the Milky Way's total stellar number. On the other hand, medium-to-low resolution spectroscopic surveys are efficient to sample millions of stars in a short period. For instance, the Large Sky Area Multi-Object Fibre Spectroscopic Telescope \citep[LAMOST;][]{CuiXQ2012} surveys have collected $R\sim1800$ spectra for more than 10 million stars in the past ten years \citep{YanHL2022}, the third data release (DR3) of the Gaia mission \citep{Gaia2016} released 220 million slitless spectra with $R\sim50$ \citep{Gaiadr32021}. Ongoing or upcoming large surveys, such as the Dark Energy Spectroscopic Instrument \citep[DESI; $2000 \lesssim R \lesssim 5500$;][]{DESI2016, DESI2022} and the Chinese Space Station Telescope \citep[CSST; $R\simeq200$;][]{ZhanH2018}, will further increase the number of stellar spectra. 

The information capacity of a spectrum is determined by not only the resolution power but also its wavelength coverage. Although low-resolution spectra loss in the sensitivity of resolving individual spectral lines, it can cover much wider wavelength range than high-resolution spectra given the same size of the detectors, leading to comparable potential of stellar label determination for a fixed exposure time \citep[e.g. see][for details]{TingYS2017a, Sandford2023}. For the data analysis of a low-resolution spectroscopic survey, the challenge is how to extract the stellar labels effectively, and in a physics rigorous manner, from the highly blended features of these spectra. This is essentially a problem of spectral modelling and optimization in high-dimensional space. 

Extensive methods of machine learning or deep learning have been implemented for deriving stellar abundances from median-to-low resolution spectra in a data-driven scheme \citep[e.g.][]{Ness2015, XiangMS2017, Ho2017, Wheeler2020, WangR2020, ZhangB2020, WangC2022, LiZ2022, LiXR2023a, LiXR2023b}. Owing to synergy between spectroscopic surveys of different resolution power, suitable cross data sets provide training labels for the machine learning models. However, while machine learning based data-driven models are flexible in high-dimensional spectral modelling, the `black box' nature of machine learning method may cause a lack of physical interpretability in the resultant labels. For instance, when estimating the value of a particular label, a data-driven model may just infer it based on astrophysical correlation with other labels, rather than predicting it rigorously based on the physical features \citep[e.g.][]{TingYS2017b}. 

To overcome this key defect, \citet{TingYS2017b} and \citet{XiangMS2019} developed the data-driven Payne ({\sc DD-Payne}), a data-driven and model-driven hybrid method that regularize the machine learning data-driven spectral modelling with differential (or gradient) spectra from physical stellar atmospheric models. While inheriting the advantages of data-driven modelling, this hybrid approach ensures the label determinations are implemented in a physics rigorous way. The {\sc DD-Payne} was applied to the analysis of LAMOST spectra, and delivered abundances of 16 individual elements for millions of stars \citep{XiangMS2019}. Examinations suggest that {\sc DD-Payne} can estimate elemental abundances from the LAMOST spectra to a precision that consistent with theoretical limit from the Cr\'amer-Rao \citep[CR;][]{Rao1945, Cramer1946} bound calculation within a factor of 2  \citep{Sandford2020, Sandford2023}. 
 
In light of this, here we apply the {\sc DD-Payne} on the early data release (EDR) of DESI \citep{DESIedr2023}). The DESI spectra cover the full optical range of 3600-9824~{\AA} at a FWHM resolution of about 1.8{\AA} \citep{DESIedr2023}. It has a wavelength-dependent resolution power ranging from 2000 to 5500, which is about twice of LAMOST. We thus expect the {\sc DD-Payne} is able to derive abundances from the spectra, reaching a higher precision limit than the latter. The DESI samples a large fraction of halo stars to a deep magnitude ($r\sim16-20$~mag), which is crucial for uncovering the  Galactic halo assembly history. Chemical abundances are important for maximizing the survey's scientific returns. 

The DESI collaboration has developed three pipelines for processing the Milky Way survey data, including the RVS, SP, and WD pipelines \citep{Cooper2023}. While the WD pipeline is focused on determining parameters for white dwarf stars, the other two pipelines determine stellar parameters of all stellar objects. The RVS pipeline can determine the radial velocities (RV), stellar atmospheric parameters ($T_{\rm eff}$, $\log\,g$, [Fe/H]) and [$\alpha$/Fe]. The SP pipeline is based on FERRE \citep{Prieto2006}, and can measure all these five parameters. In this work, we estimate 15 labels, i.e., $T_{\rm eff}$, $\log\,g$, [Fe/H], [C/Fe], [N/Fe], [Mg/Fe], [O/Fe], [Al/Fe], [Si/Fe], [Ca/Fe], [Ti/Fe], [Cr/Fe], [Mn/Fe], [Ni/Fe], and $v_{\rm mic}$, which are complements to the DESI EDR catalog.

This paper is organised as follows. Section~2 describes the DESI EDR spectra we adopted in this work. 
Section~3 introduces the method we used to determine the stellar labels. We present the main results in Section~4, and a discussion in Section~5, followed by a conclusion in Section~6.

\section{Data}
The DESI is a multi-object spectrograph wide-field survey project over 14,000 deg$^2$ \citep{DESI2016,DESI2022}. DESI can obtain fiber spectra, collected by three cameras, denoted as B (3600--5800~\AA), R (5760--7620~\AA), and Z (7520--9824~\AA), for about 5000 targets simultaneously. The resolving power of the spectra varies from $\sim2000$ at 3600~\AA, to $\sim5500$ at 9800~\AA \citep{DESIedr2023}. The DESI has five ``primary” target classes, including the Milky Way Survey program \citep[MWS;][]{Cooper2023}, bright galaxies \citep[BGS;][]{Hahn2023}, luminous red galaxies \citep[LRG;][]{ZhouR2023}, emission line galaxies \citep[ELG;][]{Raichoor2023}, and quasars \citep[QSO;][]{Chaussidon2023}. The DESI EDR targets cover a magnitude range of $16<r<20$\,mag, taking the DESI imaging surveys \citep{ZouH2017,Dey2019, Myers2023} as the photometric input catalog.

The DESI Early Data Release (EDR) was made public in 2023. The DESI EDR includes spectra from the all commissioning and Survey Validation (SV) campaigns that were taken between December 2020 and May 2021 \citep{DESIedr2023}. In this work, we use the data from the Milky Way Survey program by selecting stars from the catalog named `mwsall-pix-fuji.fits'. In total, there are 625,588 spectra for 601,782 objects with unique target ID. The catalog contains basic information, such as coordinates, observation conditions, classifications and stellar parameters derived from DESI pipelines as mentioned above in the introduction section. By inspecting the spectra, we found that some of the targets in the catalog are galaxy or quasar contamination, which can be recognised through the label ``RR\_SPECTYPE'' provided in the catalog. We therefore restrict our analysis only to those whose ``RR\_SPECTYPE'' are classified as stars, remaining 520,228 stellar spectra in our sample.  

For pre-processing, we firstly scale the spectral wavelength to the rest frame by correcting for the radial velocity of the spectra using the label `Vrad' provided by the RVS pipeline. We then normalized the spectra by dividing a pesudo-continuum. The pesudo-continuum is derived by smoothing the spectrum itself with a Gaussian kernel, 
\begin{equation}
\bar{f}(\lambda_i) = \frac{\Sigma_jf_j\omega_j(\lambda_i)}{\Sigma_j\omega_j(\lambda_i)},
\end{equation}
\begin{equation}
\omega_j(\lambda_i) = \exp\left(-\frac{(\lambda_j-\lambda_i)^2}{2*L^2}\right), L = 20\AA.
\end{equation}
Here we adopt a Gaussian kernel width $L$ of 20{\AA}. 
In this step, we mask the center of strong Balmer lines and diffuse interstellar bands (DIBs). Also, for bad pixels with `NaN'  values in the flux or unreliable large flux invariance, we set the flux errors of these pixel as infinity (999999. in action). 

\section{Method} 
\label{section}
\subsection{ The Data-Driven Payne}
The {\sc DD-Payne} is a data-driven spectral modelling method regularized by differential spectra from physical stellar atmospheric models. It was initiated by \citet{TingYS2017a}, and well developed and applied to the LAMOST survey spectra by \citet{XiangMS2019}. In brief, the {\sc DD-Payne} models the stellar spectra with a multi-layer perception (MLP) neural network, which is trained on a set of the survey spectra with well-known labels. The key philosophy is that in the training process it adopts a loss function regularized by differential spectra from stellar atmospheric models, as illustrates in the second term of the Equation below,
\begin{equation}
\begin{aligned}
&  \mathcal{L}(\{f_{\rm obs}(\lambda)\} | \mathbf{w},\mathbf{b}) =  \frac{1}{N_s}\sum_{i=1}^{N_s} 
         \frac{(f(\lambda | \boldsymbol{\ell}_{{\rm obs},i}) - f_{{\rm obs},i}(\lambda))^2}{\sigma^2_{{\rm obs},i}(\lambda)}  \\
 &   +\sum_{j=1}^{N_r} \mathbf{D}_{\rm scale} \cdot \sum_{k=1}^{N_l} | f^{\prime}(\lambda | \boldsymbol{\ell}_{\rm ref}) - f^{\prime}_{\rm {physical}}(\lambda | \boldsymbol{\ell}_{\rm ref})|
\end{aligned} 
\end{equation}
where $f^{\prime}_{\rm {physical}}$ is the gradient spectra by stellar physical models, $D_{scale}$ is a vector parameters to determine the weight of each label, the number of training spectra, reference stars, and stellar labels, are named as $N_{s}$, $N_{r}$, $N_{l}$, respectively. Similar to \citet{XiangMS2019}, the $\mathbf{w}$ and $\mathbf{b}$ are coefficient arrays to be optimized of the MLP neural network.

{\sc DD-Payne} is able to extract many stellar labels simultaneously from a spectrum in a physics-rigorous manner. The key features of {\sc DD-Payne} include several aspects. First, it can deal with high-dimensional ($>20$) spectra modelling through a flexible neural network algorithm. The model can properly predict not only the spectral flux but also the gradient spectra by imposing regularization of differential spectra from stellar atmospheric models in the training process. Second, it employs the all-pixel fitting strategy to make full use of the spectral features. This is vital for both improving the robustness of the fitting and breaking up degeneracy among various labels of the low-resolution spectra. Third, it fits all the labels at the same time, which is necessary to reach a global minimum in the optimization. Finally, adopting a data-driven scheme ensures the resultant labels having the same scale with those of the training sample, making it as a self consistent spectral modelling. 
\begin{figure}[htb!]
\centering
\includegraphics[width=0.47\textwidth]{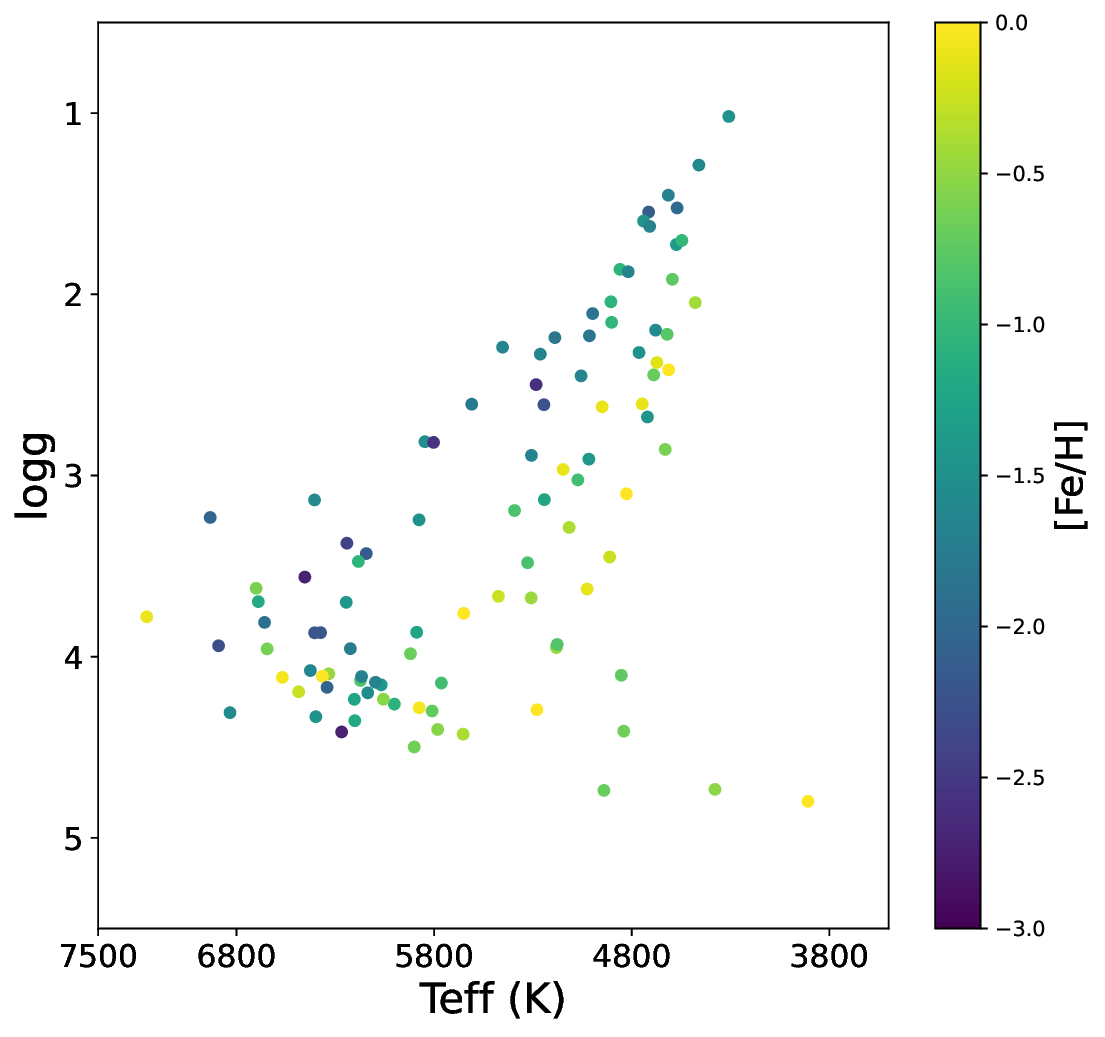}
\caption{$T_{\rm eff}$--$\log\,g$ distribution of the stellar grids for which the differential spectra are used to regularize the {\sc DD-Payne} in the current work. The $T_{\rm eff}$, $\log\,g$, and $\feh$ of these grids are drawn from the MIST isochrones \citep{Choi2016}. In total, there are 101 grids spanning an $\feh$ from $-3$ to 0.5. For each grid, 15 spectra are computed with the Kurucz ATLAS12 atmospheric model to provide the physical differential spectra for 15 labels.   
\label{figkurucz}}
\end{figure}

In this work, we make some further improvements and tailors to the method on top of these works. The major update is that we significantly increase the number of gradient spectra, computed with the Kurucz atmospheric models \citep{Kurucz1970, Kurucz1993, Kurucz2005}, for the regularization. We now have 101 stellar model grids, while \citet{XiangMS2019}  used 16 only. Increasing model grids provides better constraints of the {\sc DD-Payne} differential spectra across a wider range of parameter space from $\feh=-3.0$ to $\feh=0.5$, as shown in Fig.\,\ref{figkurucz}.  All these spectra are tailored to fit the resolution of DESI. We still keep the same architecture of the neural network model as \citet{XiangMS2019}, i.e., adopting a 2-layer network, but we empirically increased the number of neurons from 10 to 40.
\subsection{The training set}
\begin{figure*}[htb!]
\centering
\includegraphics[width=0.97\textwidth]{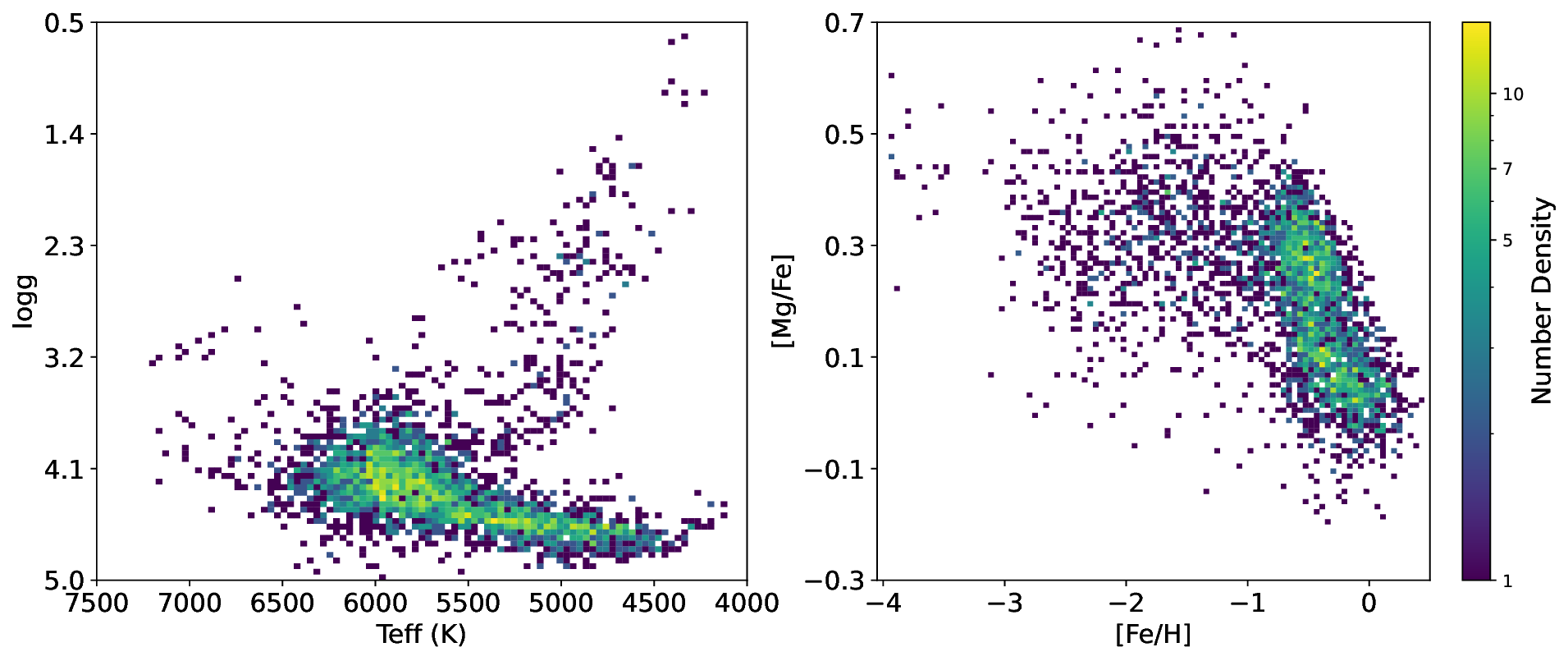}
\caption{Stellar density distribution of the training stars in the $T_{\rm eff}$--$\log\,g$ diagram ($left$) and the [Fe/H]--[Mg/Fe] plane ($right$). The labels of the training sample stars are derived from either APOGEE DR17 or LAMOST DR9 {\sc DD-Payne}.
\label{fig1}}
\end{figure*}
Ideally, labels of the training set are ought to be derived from high-precision measurements, e.g., from high-resolution spectroscopy. At first, we cross-match the DESI EDR catalog with the APOGEE DR17 \citep{Abdurrouf2022}. However, there are only 523 common stars between these two catalogs, but only 267 stars have a spectral signal-to-noise ratio ($S/N$) higher than 20 in the blue band and have available stellar elemental abundance measurements in APOGEE DR17. This is a too small data set to be taken as a training set. 

As a comprise, we opt to adopt the LAMOST stellar labels as the training labels. LAMOST DR9 has been released in 2023, and contains 10,907,516 low-resolution ($R\sim1800$) spectra. 
Zhang et al. (in prep) have determined the stellar parameters and abundances with the {\sc DD-Payne} by taking common stars with APOGEE DR17 as the training set. The LAMOST metal-poor sample of \citet{LiHN2018} is also added as the training set to expand the metallicity coverage of reliable measurements from LAMOST DR9 down to $\feh\simeq-4$.
A cross-match between DESI EDR and LAMOST DR9 results a set of 3181 common stars with $S/N > 20$ for both the DESI and LAMOST spectra.

Finally, the training sample in this work contains 3627 stars from a combination of two samples, i.e., common stars between DESI and LAMOST DR9, and common stars between DESI EDR and APOGEE DR17.
Fig.~\ref{fig1} shows the distribution of the training stars in the $T_{\rm eff}$--$\log\,g$ diagram. The sample spreads a large parameter space. The majority of them are main-sequence and turn-off stars. 
While the majority of the stars are metal-rich ($\feh>-0.8$) disk stars, there is a sample of metal-poor stars, which has metallicity down to $\feh\simeq-4$. 
\subsection{Training the DD-Payne model}
\begin{figure}[htb!]
\centering
\vspace{1.em}
\includegraphics[width=0.48\textwidth]{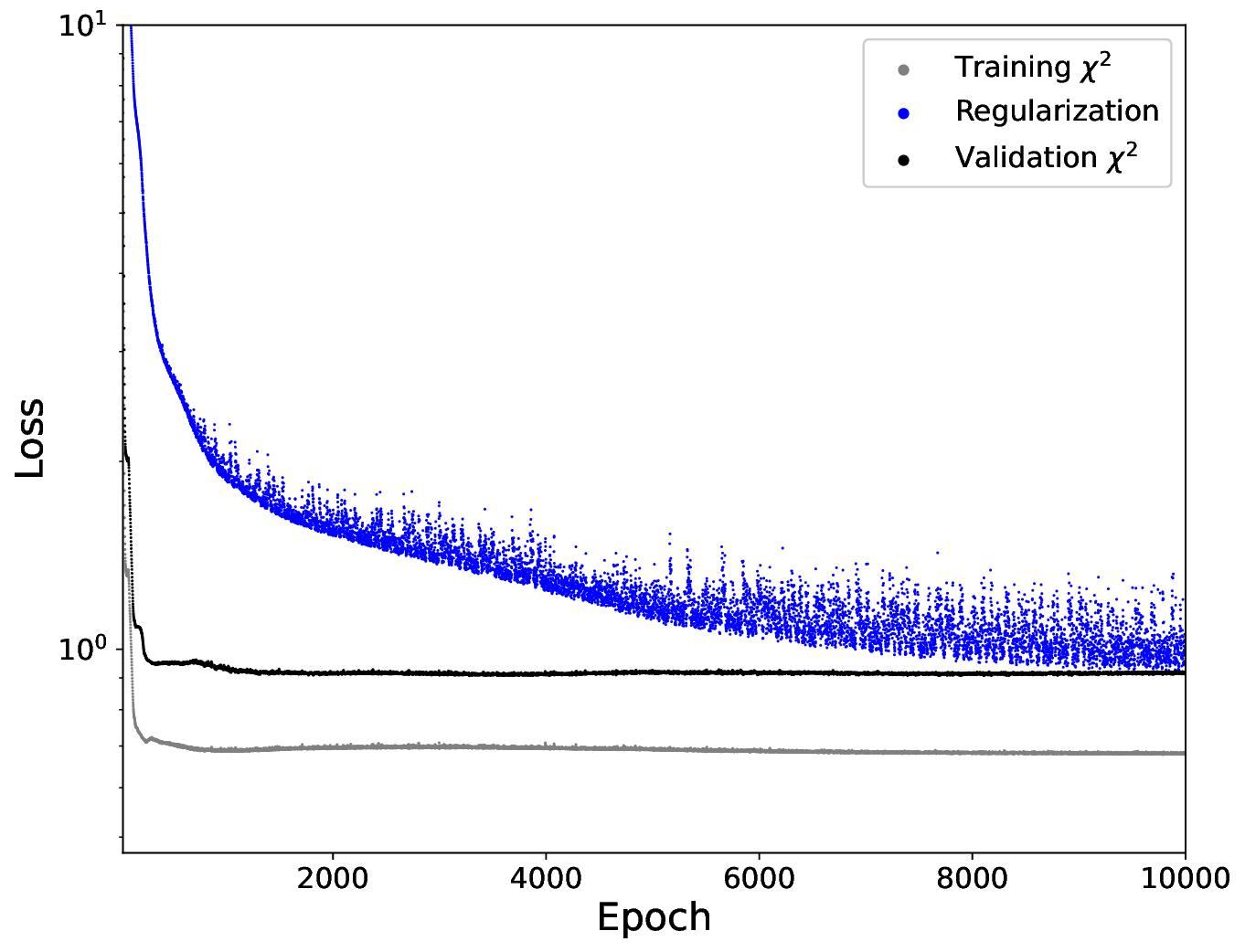}
\vspace{0.5em}
\caption{Loss values as a function of epoch of the training process. This figure shows the fifteenth pixel of spectrum with the our training sample. As shown in the legend, the grey curve shows the loss value contributed by spectral flux of the training set, and the blue one shows the contribution of regularization term by differential spectra. The black curve shows the loss value contributed by spectral flux of the validation sample.  
\label{fig2}}
\end{figure}
\begin{figure*}[htb!]
\centering
\includegraphics[width=0.97\textwidth]{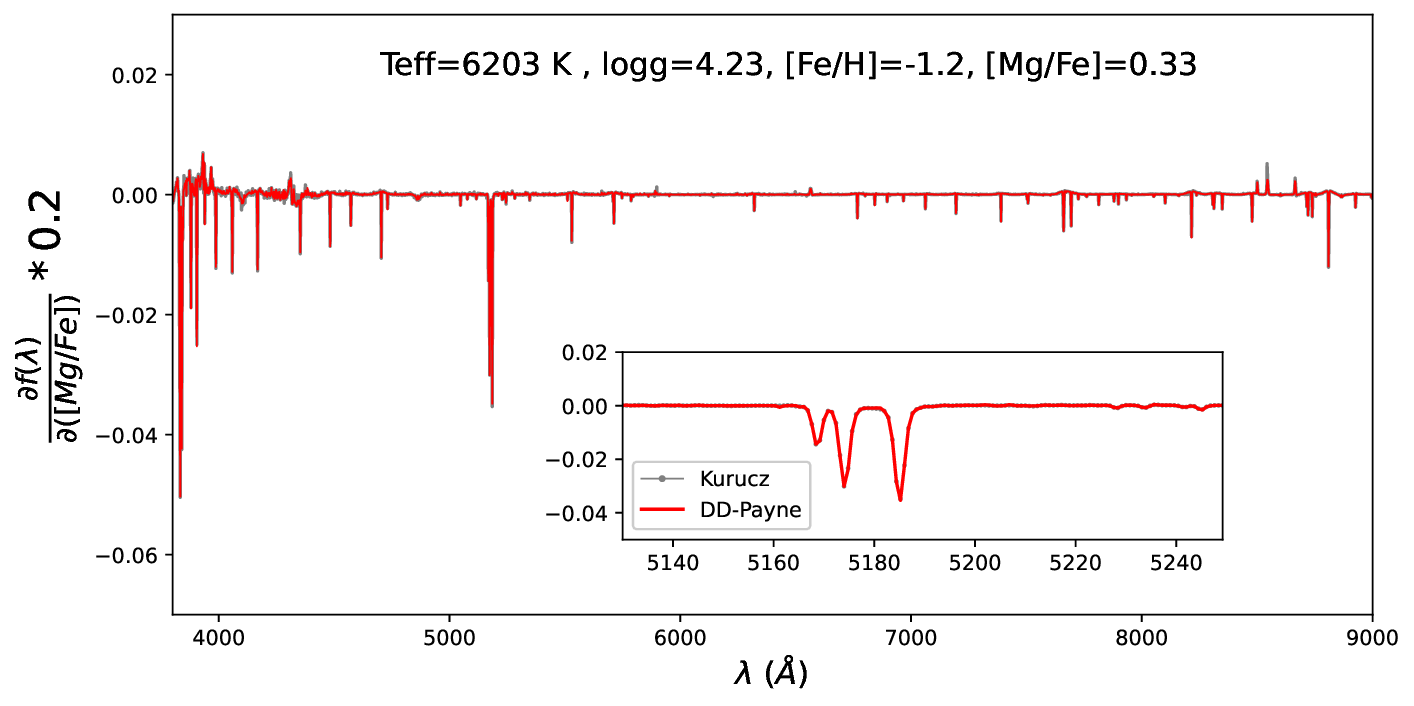}
\caption{Comparison of differential spectrum for Mg between the {\sc DD-Payne} prediction (red) and the Kurucz model calculation (grey) for a fiducial star, whose $T_{\rm eff}$, $\log\,g$, [Fe/H], and [Mg/Fe] values are labeled in the figure. The zoomed-in plot highlights the Mg\,{\sc i b} line region. The excellent match between the {\sc DD-Payne} and the Kurucz model in the differential spectrum ensures that the former determine Mg abundance using physical features. These features are either directly the Mg lines or lines of other species affected by the opacity change due to the increase of Mg abundance.  
\label{fig3}}
\end{figure*}
The neural network spectral model training was implemented in the $Pytorch$ environment. The training sample is divided into batches with a size of 512. We train the 2-layer neural network pixel by pixel, and the neural network of each pixel has 40 neurons. In the training process, we allow the learning rate vary from 0.01 at the beginning to 0.0001 at the end. The training process is stopped after 10,000 iteration epochs. For the regularization of the gradient spectra, we assign weights to different labels to account for their different strengths. 

As an example, Fig.\,\ref{fig2} shows the loss function, for a typical pixel, as a function of the iteration epoch.
It drops quickly and reaches a plateau after about a few hundred iterations. However, it takes much more iterations for the converge of the regularization term, i.e., to build a correct model for the first derivative. This demonstrates the necessarity to impose the regularization of gradient spectra in order to build a physical sensible data-driven model. Fig.\,\ref{fig3} further shows a comparison of the gradient spectrum for [Mg/Fe] between the {\sc DD-Payne} model and the Kurucz model. The consistency illustrates that the {\sc DD-Payne} reproduces the first derivation of the spectra model well, guaranteeing a physical rigorous determination of the stellar labels.

\section{Results}
Our implementation of the {\sc DD-Payne} on the DESI EDR data set results in a catalog of stellar parameters ($T_{\rm eff}$, $\log\,g$, $v_{\rm mic}$) and abundances of 12 elements (C, N, O, Mg, Al, Si, Ca, Ti, Cr, Mn, Fe, Ni) from 629,934 spectra. Among them, 520,228 spectra are classified as `STAR' through the `RR\_SPECTYPE' in the DESI catalog, while others are classified as `QSO' or `GALAXY'. To identify outliers, we inspect the spectral fitting and introduce a quality flag ${\rm qflag}\_{\chi^2}$, 
\begin{align}
     {\rm qflag}\_\chi^2 = \frac{(\chi^2 - \chi^2_{\rm median})}{\chi^2_{\rm mad}}.
\end{align}
Here the $\chi_{\rm median}^2$ and $\chi_{\rm mad}^2$ are the median and mean absolute deviation from the median $\chi^2$ values for stars of similar spectral $S/N$s. In the following analysis, we only focus on objects that are classified as `STAR' in DESI catalog and have good fits (${\rm qflag}\_\chi^2<3.0$) in our work, to eliminate contamination of galaxy/quasar and stellar outliers, respectively. 

In addition to the stellar abundances, we also provide an estimate of the extinction and distance, as well as orbital parameters for the sample stars (Sect.\,4.3).

\subsection{Basic stellar atmospheric parameters}
\begin{figure*}[htb!]
\centering
\includegraphics[width=0.97\textwidth]{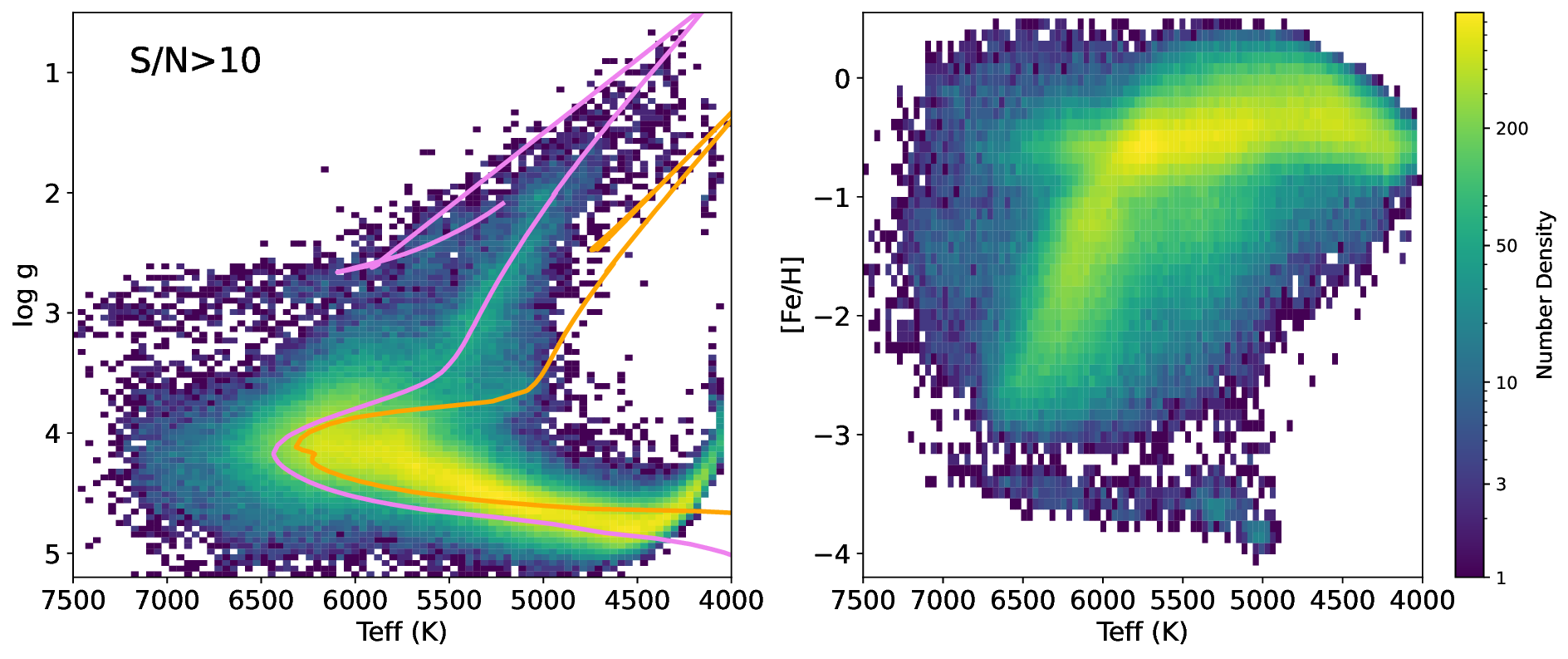}
\vspace{0.5em}
\caption{Stellar number density distributions of the DESI EDR stars with blue band spectra $S/N>10$ in the $T_{\rm eff}$--$\log~g$ diagram (left) and $T_{\rm eff}$--[Fe/H] plane (right). These stellar parameters in this figure are measured by {\sc DD-Payne} method. The $S/N>10$ cut remains 249,877 stars shown in each panel. The violet line in the left panel shows the stellar isochrone for an age of 12~Gyr and [Fe/H] of $-1.5$, extracted from the PARSEC database \citep{Bressan2012}, while the orange line shows the isochrone for an age of 4.0~Gyr and [Fe/H] of 0.0 (solar metallicity).
\label{fig4}}
\end{figure*}
\begin{figure*}[htb!]
\centering   
\subfigure
{
	\begin{minipage}{0.3\linewidth}
	\centering  
	\includegraphics[width=0.98\columnwidth]{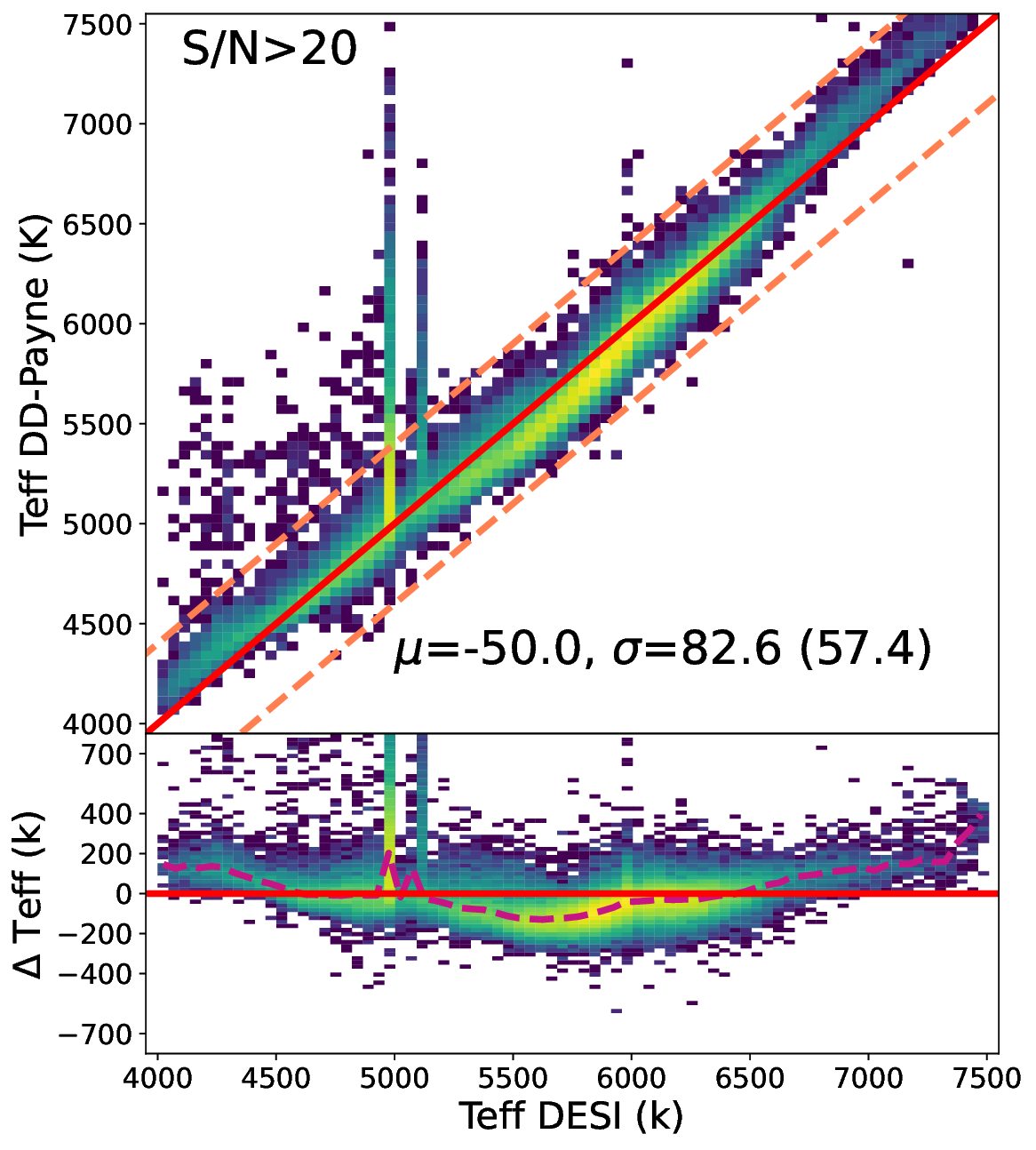} 
	\end{minipage}
} 
\subfigure
{
	\begin{minipage}{0.29\linewidth}
	\centering  
	\includegraphics[width=0.97\columnwidth]{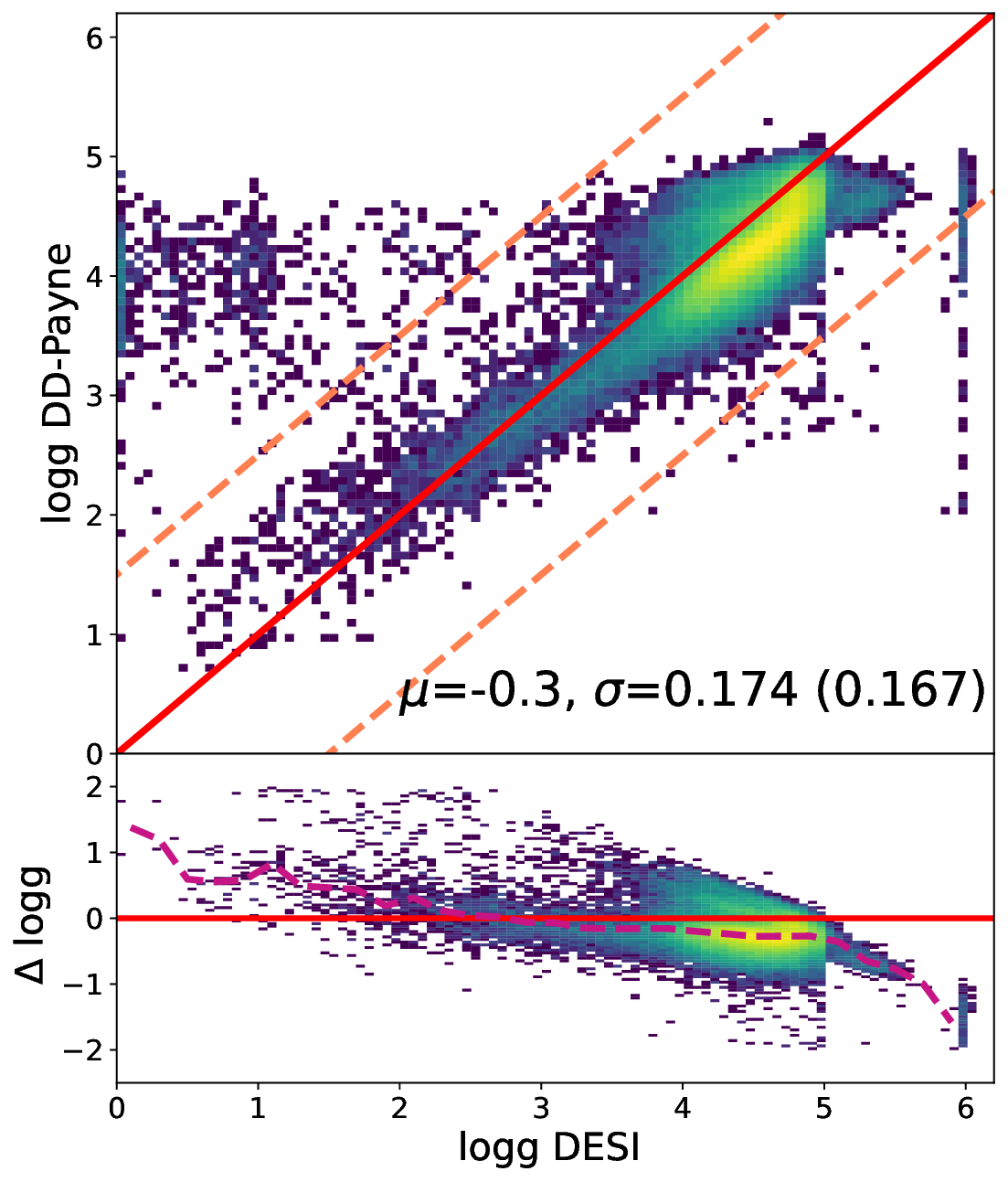} 
	\end{minipage}
} 
\subfigure
{
	\begin{minipage}{0.3\linewidth}
	\centering  
	\includegraphics[width=0.97\columnwidth]{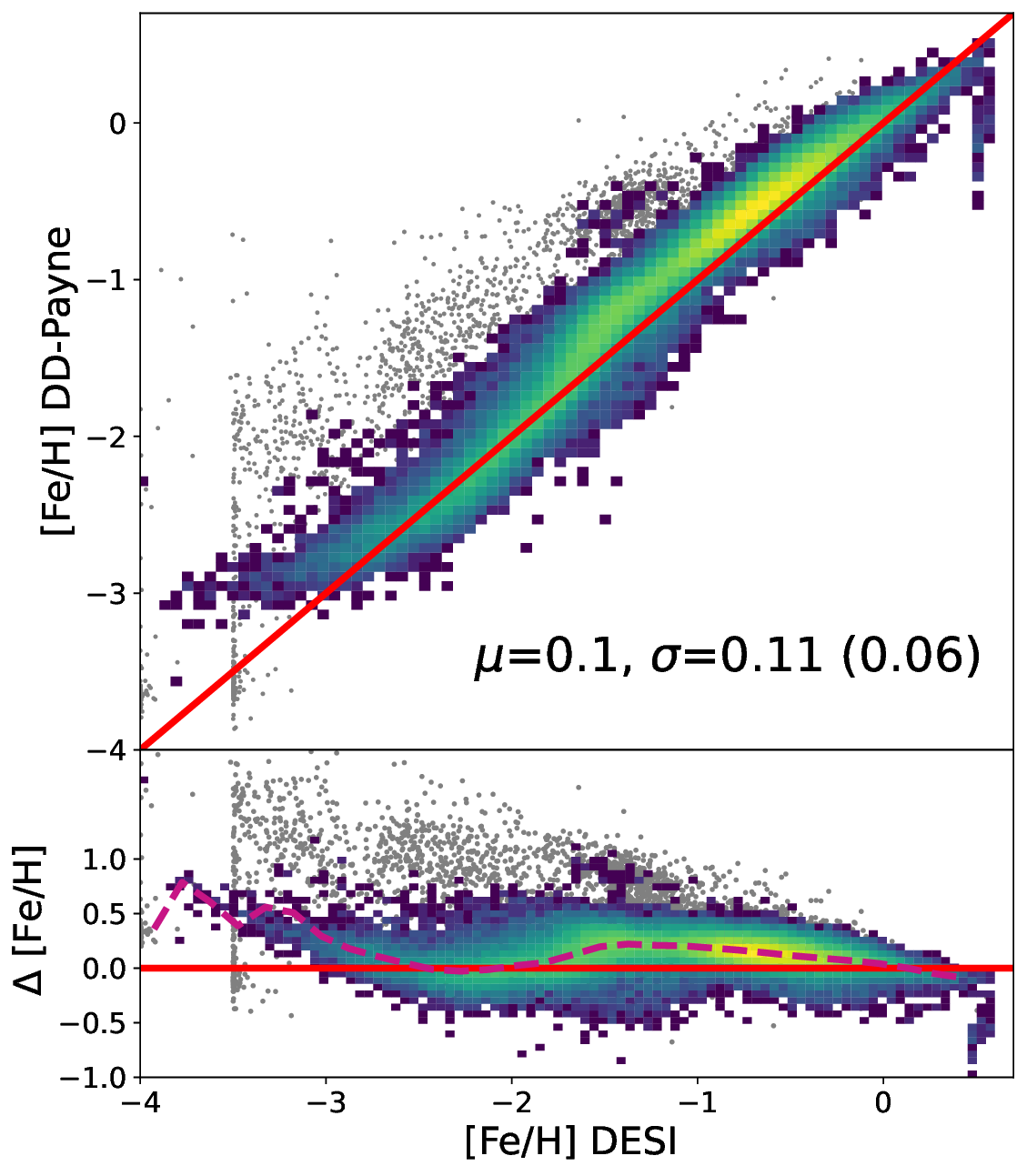} 
	\end{minipage}
} 
\caption{The comparison of the basic atomospheric stellar parameters (from the left to the right are $T_{\rm eff}$, $\log~g$, and [Fe/H]) between the {\sc DD-Payne} estimates and the DESI EDR catalog values by SP pipelines. Color represents the number density of stars. Here we only show stars with $S/N>20$ and good quality flag in both DESI EDR and our measurements. Solid lines in the upper panels show the 1:1 line of X and Y axes. The dotted lines in the upper-middle panel indicate a $T_{\rm eff}$ offset of $\pm400$~K relative to the 1:1 line. Similarly, the dotted lines in the middle panel indicate a $\log~g$ offset of $\pm1.5$~dex relative to the 1:1 line. In the right panels, stars outside the dotted lines in $T_{\rm eff}$ (upper-left panel) and $\log~g$ (upper-middle) panels are shown in grey. 
The mean and dispersion of the differences between {\sc DD-Payne} and the DESI EDR parameters are marked in the figure. The number in the bracket shows the dispersion after substracting the systematic trend in the difference as indicated by the violet line in the bottom panels.    
\label{fig5}}
\end{figure*}
Here we characterize the distributions of basic stellar atmospheric parameters, including $T_{\rm eff}$, $\log~g$, and metallicity [Fe/H]. The left panel of Fig.~\ref{fig4} shows the stellar density distribution in $T_{\rm eff}$--$\log\,g$ (Kiel) diagram for stars with $S/N>10$ in the blue band and a fitting $\chi^2$ value smaller than 3, which leaves 249,877 stars in the sample. The sample spans a temperature range of 4000-7500\,K, as most of the sample stars are F/G/K type. A few stars in the sample may have true temperatures beyond this range, e.g., cool M or hot, OBA type stars. However, due to limitation of the training set (Fig.~\ref{fig1}), our method currently can't properly determine the parameters for them. 
 
The sample stars spread different evolutionary phases, including main-sequence dwarfs, turn-off/subgiants, red giant, and blue and red horizontal branch (BHB/RHB). Most of them ($>90\%$) are main-sequence dwarfs or turn-off stars, while the red giant branch stars only constitute less than 5\% of the sample. It is noticeable that the horizontal branch can be well recovered, suggesting that the {\sc DD-Payne} works well in determining $\log~g$ for the stars at the border of the training set (Fig.~\ref{fig1}). There is a blob of stars in the range of $5000<T_{\rm eff}<5500$\,K and $3.5<\log~g<4$, which shows cut off at cooler temperature end with lower $\log~g$. This turns out to be a consequence of the DESI target selection function: the DESI EDR mainly targets stars with $16<r<20$\,mag towards the Galactic halo ($|b|\gtrsim20^\circ$), so that metal-rich thus cool disk stars in the red giant branch were omitted by the bright-end magnitude limit. Finally, note that for dwarf stars at the low temperature border ($T_{\rm eff}<4500$~K), the $\log~g$ shows an arbitrary rising with decreasing temperature. This is an artifact inherited from the training sample, which is further magnified by extrapolation of the {\sc DD-Payne}.    

The right panel of Fig.~\ref{fig4} shows the stellar density distribution in $T_{\rm eff}$--$\feh$ plane. The sample stars span a metallicity range of $-4.0$ -- $0.5$\,dex. The metal-rich ($\feh>-1$) part is dominated by disk populations, while the metal-poor ($\feh<-1$) part is expected to be dominated by halo populations. For the disk stars, the bulk metallicity occur at $\feh\simeq-0.5$, indicating they are dominated by thick disk stars rather than thin disk stars, which peaks at $\feh\simeq-0.1$ \citep[e.g.][]{XiangMS2019, WangC2019}.
Since the DESI EDR targets relatively faint objects at high Galactic latitudes, it tends to omit local, thin disk objects. There is a substantial set ($\sim30\%$) of metal-poor halo stars down to a metallicity of $\feh\simeq-4$. It shows a significantly higher sampling rate for metal-poor stars compared to other spectroscopic surveys (Sect.~5), which is owing to the deeper magnitude range of the DESI survey. Stars with $\feh<-0.8$ exhibit a clear trend that the more metal-poor ones have higher temperatures. As these metal-poor stars are mostly main-sequence turn-offs, this trend is just a manifestation that they share similar (old) ages \citep[e.g.][]{XiangMS2022b}. At the metal-poor end ($\feh<-3$), the stars exhibit a weird distribution that is separated from the more metal-rich ones. We have inspected the spectra of stars in these extremely metal-poor end, and found most of them have relatively low spectral $S/N$s ($\simeq10$) and exhibit board Balmer absorption lines. We suspect they would be either white dwarfs or hot sub-dwarfs, whose parameters are erroneously estimated due to limitation of the training set. The $\feh$ estimates for these stars need to be used with great cautious.

Fig.~\ref{fig5} shows a comparison between our parameter estimates and the DESI EDR official release from DESI SP pipelines \citep{DESIedr2023}. The effective temperatures exhibits a good one-to-one consistency, the differences have a dispersion of only 56.1~K, reflecting the good robustness of both methods for parameter determination. However, there is an average systematic offset of 46.7~K, which also depends on the effective temperature itself. These systematics are consequences of different temperature scales adopted: the DESI EDR scale is spectroscopic temperature  determined using the Kurucz synthetic spectra, while our scale is inherited on the LAMOST DR9 results calibrated to the infrared flux method of \citep{Gonza2009} (see Zhang et al. in prep. for details). Beyond the overall good consistency, the left panel of Fig.~\ref{fig5} exhibits two vertical spikes, which are likely artifacts in the DESI EDR estimates, whose positions suggest that they are possibly related to template grids adopted by DESI pipeline. 

The middle panel of Fig.~\ref{fig5} illustrates that the $\log~g$ exhibits also good overall consistency between our estimates and those of DESI SP pipeline values. The dispersion of their differences is only 0.17\,dex. However, the dwarfs exhibit a systematic difference of 0.3~dex. Generally, it is extremely difficult to build accurate synthetic spectra in the full optical wavelength range due to imperfections in (atomic and molecular) line lists and opacity, etc. Therefore it is not surprising that the $\log~g$ derived by fitting the full, low-resolution (blended) spectra suffers some moderate systematic differences. This is one of the advantages for applying data-driven approaches to low-resolution spectra. The training sample of {\sc DD-Payne} method adopts $\log~g$ values inherited from the APOGEE high-resolution spectroscopy (Zhang et al. in prep.), so that we expect our estimates are more accurate than the DESI EDR values. Beyond this, there is a group of stars that our estimates suggest dwarfs ($\log~g>3$) while DESI EDR suggest giants ($\log~g<2$). These stars are shown as the vertical spikes in the $T_{\rm eff}$ comparison plot (the left panel of Fig.\,\ref{fig5}), which are likely to be estimated improperly by SP pipeline. 

The right panel of Fig.~\ref{fig5} illustrates a good consistency between our [Fe/H] estimates and the DESI EDR pipeline values, with [Fe/H] values down to $-3.0$. The overall dispersion of the [Fe/H] differences is only 0.1~dex (0.06~dex after a systematic trend correction). However, there is a significant [Fe/H]-dependent systematic difference. The detailed reason for such difference is not entirely clear. Generally, for estimating parameters from low-resolution, blended spectra, there is strong covariance (correlation) among $T_{\rm eff}$, $\log~g$ and [Fe/H] \citep[e.g.,][]{TingYS2017b, XiangMS2019, XiangMS2022}. As a result, an over-estimate of the $\log~g$ value for using imperfect models often comes along with an underestimate of [Fe/H] value. This means the [Fe/H] values may affected by some systematic errors of $\log~g$ estimates of the DESI EDR SP pipelines. On the other hand, our estimates of [Fe/H] values are also affected by complications of the training sets: the [Fe/H] values of LAMOST are derived using the combination of two separate training sets, the APOGEE DR17 \citep{Abdurrouf2022} for the relatively metal-rich part ($\feh\gtrsim-1.5$) and the LAMOST metal-poor star sample ($\feh<-1.7$) of \citet{LiHN2018}. This may contribute to the [Fe/H]-dependent trend in the figure.    

\subsection{Stellar abundances} 
\begin{figure*}[htb!]
\centering
\subfigure
{
	\begin{minipage}{0.97\linewidth}
	\centering  
	\includegraphics[width=0.97\columnwidth]{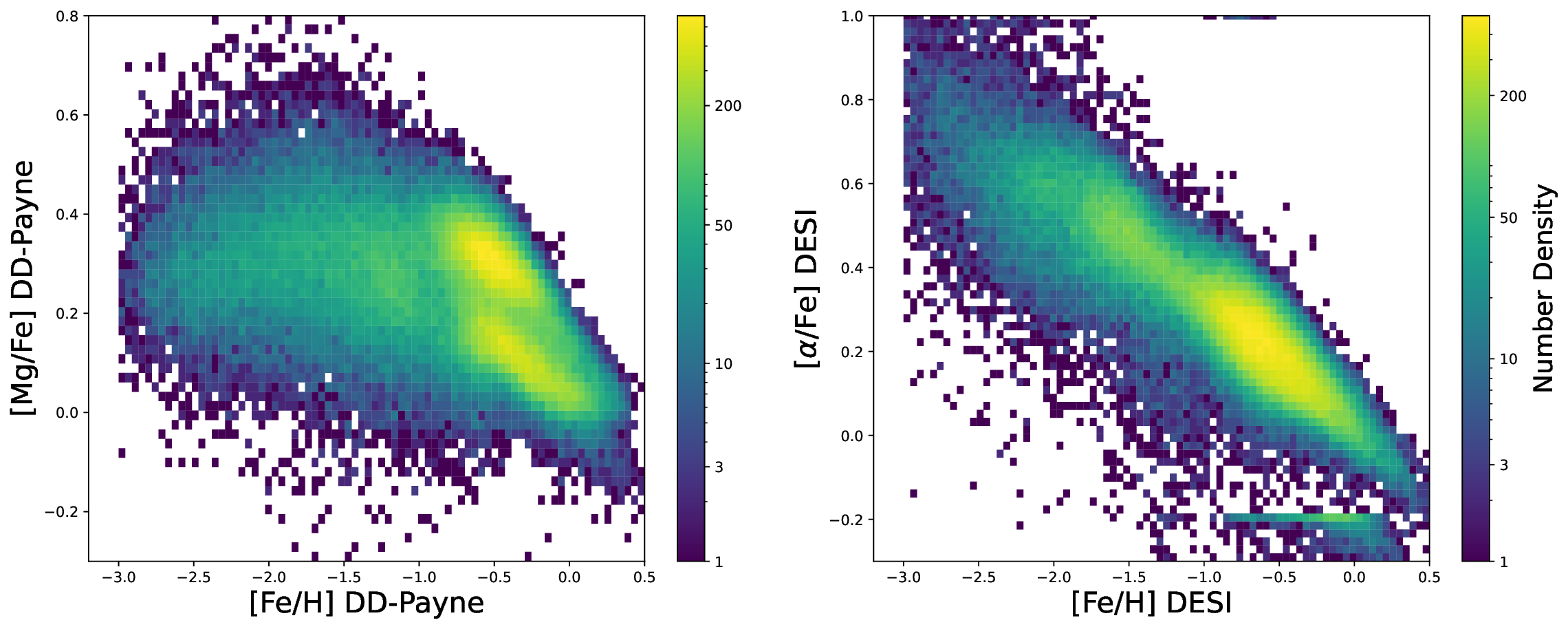}
	\end{minipage}
} 
\subfigure
{
	\begin{minipage}{0.97\linewidth}
	\centering     
	\includegraphics[width=0.95\columnwidth]{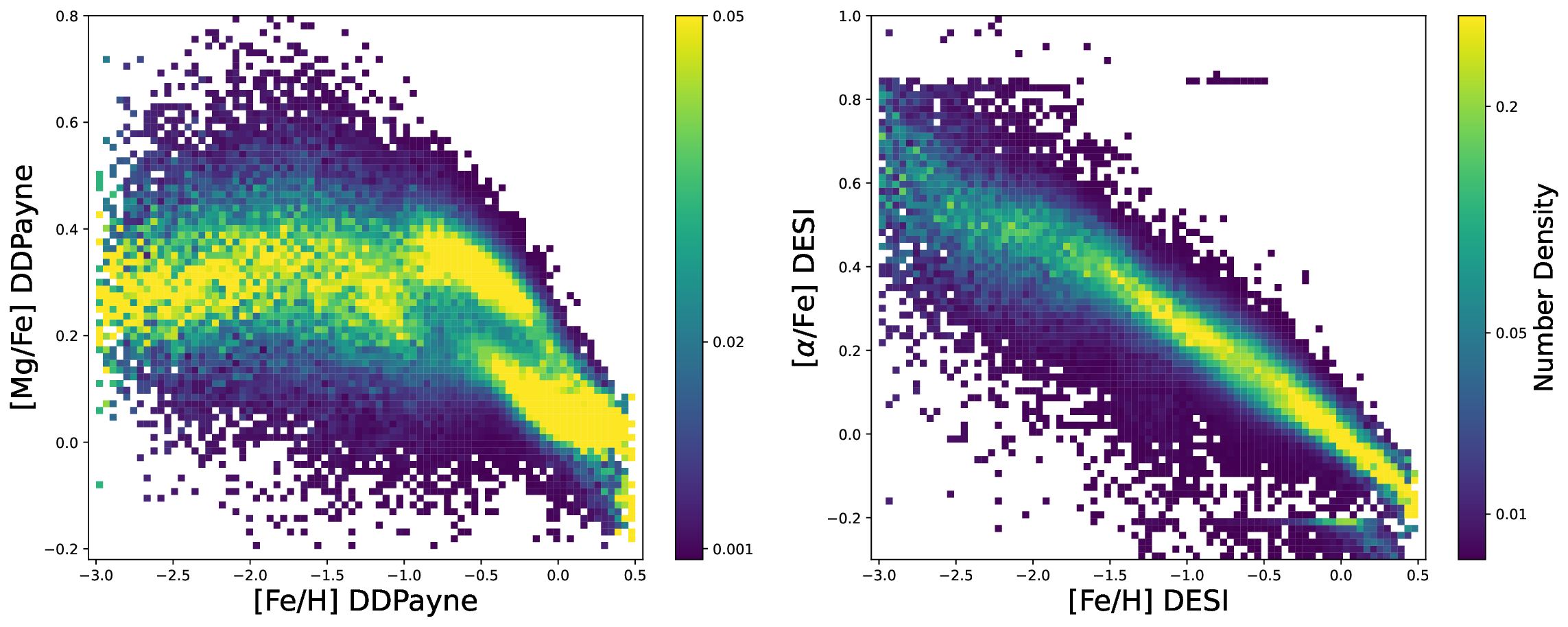}  
	\end{minipage}
}
\caption{Stellar number density distribution in the [Fe/H]-[Mg/Fe] plane for the {\sc DD-Payne} results (left two panels) and distribution in the [Fe/H]-[$\alpha$/Fe] plane for the DESI EDR results by SP pipeline (left two panels). The bottom panels are a column-normalized version of the top panels. Only stars with blue band spectral $S/N>20$ are shown in the figure.
\label{fig6}}
\end{figure*}

The upper-left panels of Fig.~\ref{fig6} shows the stellar distribution in the [Mg/Fe]-[Fe/H] plane. It clearly presents a bimodal distribution at the metal-rich ($\feh>-1$) end, which corresponds to the well-known high-$\alpha$ and low-$\alpha$ disk populations \citep[e.g.][]{Hayden2014}. At the metal-poor end, the distribution is more dispersed, reflecting both the complex abundances of halo populations and the larger measurement errors. 
Since most of the sample stars are still metal-rich ([Fe/H]$>-1.0$), we look into the [Mg/Fe] distribution of the stars with similar metallicity.
As shown in bottom left panel of Fig.~\ref{fig6}, there reveals a sequence with [Mg/Fe] values decreasing from $\sim0.3$~dex at $\feh=-1.5$ to $\sim0.0$~dex at $\feh\gtrsim-1$. This sequence is constituted mostly by remnant stars of the Gaia-Enceladus-Sausage (GES) galaxy \citep{Belokurov2018, Helmi2018}, which is verified as the dominated halo population \citep{Helmi2020}. The metal-poor end of $\feh\lesssim-1.5$ exhibits a prominent feature, which has a near constant [Mg/Fe]. These populations might be a part (or related to) an in-situ component of the halo, the so-called {\em Aurora} \citep{Belokurov2022}.  

As a comparison, the right panels show the DESI EDR [$\alpha$/Fe]-[Fe/H] map. The double sequences of the disk can hardly be discernible from the [$\alpha$/Fe], and the GES is also hardly discernible from high-$\alpha$ disk at all metallicity. We believe this is because the overall $\alpha$ abundances reflect the trend of Si, Ca, Ti abundances, which do not show clearly distinct values among the different [Mg/Fe] sequences, as seen in Fig.~\ref{fig6}. Exercises suggest that if we average abundances of all the $\alpha$-elements O, Mg, Si, Ca, Ti, we would obtain an [$\alpha$/Fe] value in agreement with the DESI estimates. This highlights the importance to derive individual elemental abundances than the overall $\alpha$ abundance from the spectra.

\begin{figure*}[htb!]
\centering
\includegraphics[width=0.97\textwidth]{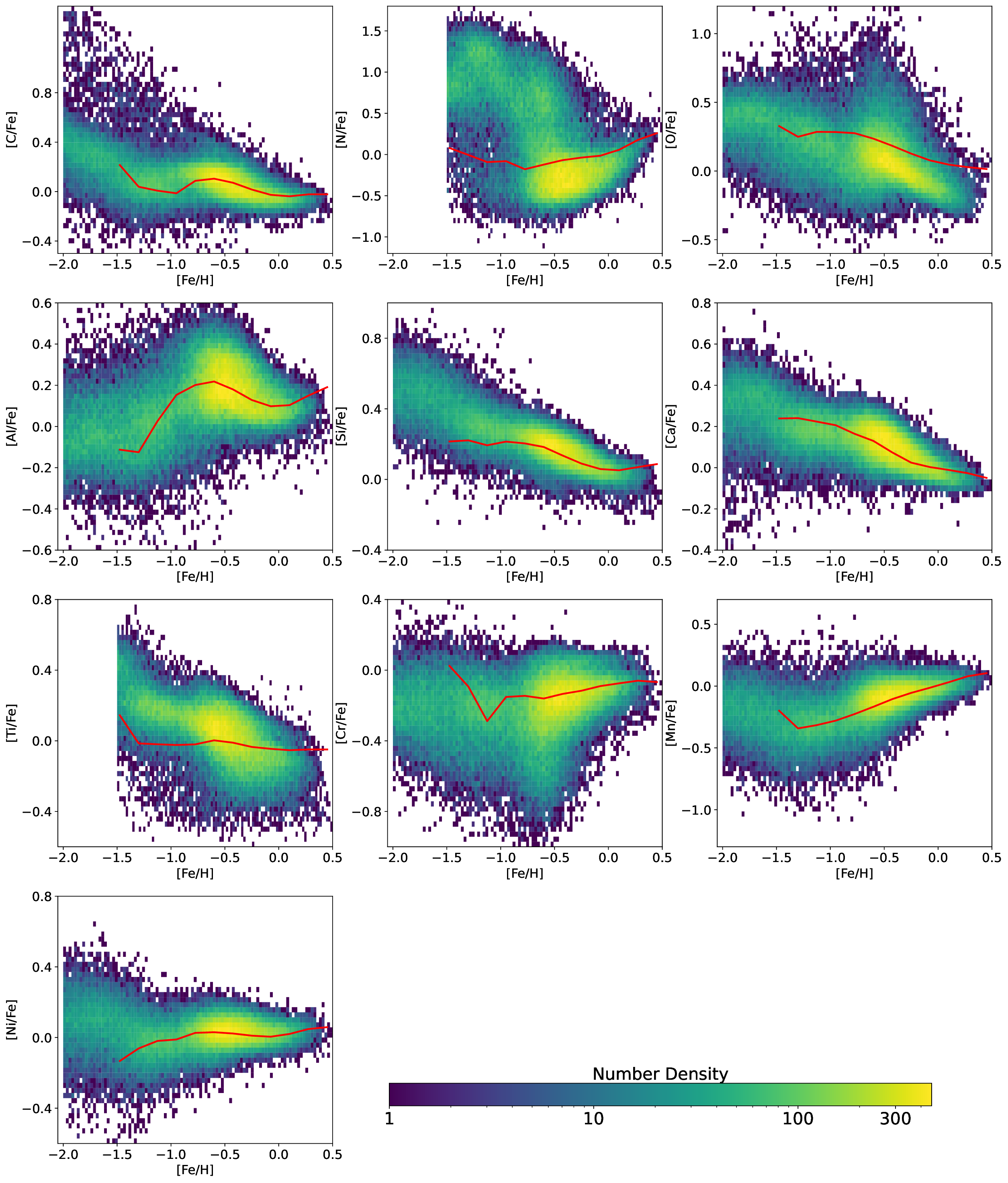}
\caption{Stellar number density distributions in the [X/Fe]-[Fe/H] abundance planes. Owing to limitation of the training sample, different cuts at lower [Fe/H] ends are made for different elements. For N and Ti, a cut of $\feh=-1.5$ is taken, while a cut of [Fe/H] $=-2.0$ is taken for C, O, Al, Si, Ca, Cr, Mn, and Ni. For a guidance, the red line shows the abundance trend of dwarf stars in the APOGEE DR17. For the majority of the elements, the stellar abundances from DESI spectra are consistent well with the APOGEE DR17 trend.
\label{fig7}}
\end{figure*}

The [X/Fe]-[Fe/H] map for other elements are shown in Fig.~\ref{fig7}, along with an overall abundance trend of the APOGEE dwarf stars. The [C/Fe], [O/Fe], and [Al/Fe] show distinct trends between the disk ([Fe/H]$>-1$) and the halo populations ($\feh<-1$), which are well consistent with the APOGEE high-resolution results. The [N/Fe] shows slight different trend to APOGEE, with unclear reason.  The overall trend of [Si/Fe], [Ca/Fe], [Ti/Fe] are consistent with the APOGEE, but different to that of the [Mg/Fe] as shown in Fig.~\ref{fig6}. This is likely due to different synthesize paths: Mg is almost exclusively produced by Type II supernovae, while a portion of Si, Ca and Ti are produced by Type Ia supernovae \citep{Kobayashi2020}. Iron-peak elements Cr and Ni exhibit an abundance ratio close to zero with small scatters, which are consistent with APOGEE as well as other previous results \citep[e.g.][]{XiangMS2019}. The [Mn/Fe], however, shows a slight positive trend with [Fe/H] in the metallicity range of [-1.5, 0.5], which is again consistent with the APOGEE and other studies. There is small portion of stars with exceptionally low [Cr/Fe] at an [Fe/H] of $\sim-0.6$. This is found to be an artifact inherited from training set. 

Note that due to limitation of the training sample, stars in the very metal-poor regimes are hard to have accurate abundance determinations. We therefore have opted to restrict our analysis to stars above a $\feh$ threshold, whose value varies among different labels. For N and Ti, the abundances are cut at $\feh=-1.5$, as we do not have reliable training labels below this threshold, while for other elements the abundances are cut at $\feh=-2$. We expect that there will be better training sets in future releases, thus one can push these thresholds to lower $\feh$ values.  

\begin{figure*}[htb!]
\centering
\vspace{1.em}
\includegraphics[width=0.97\textwidth]{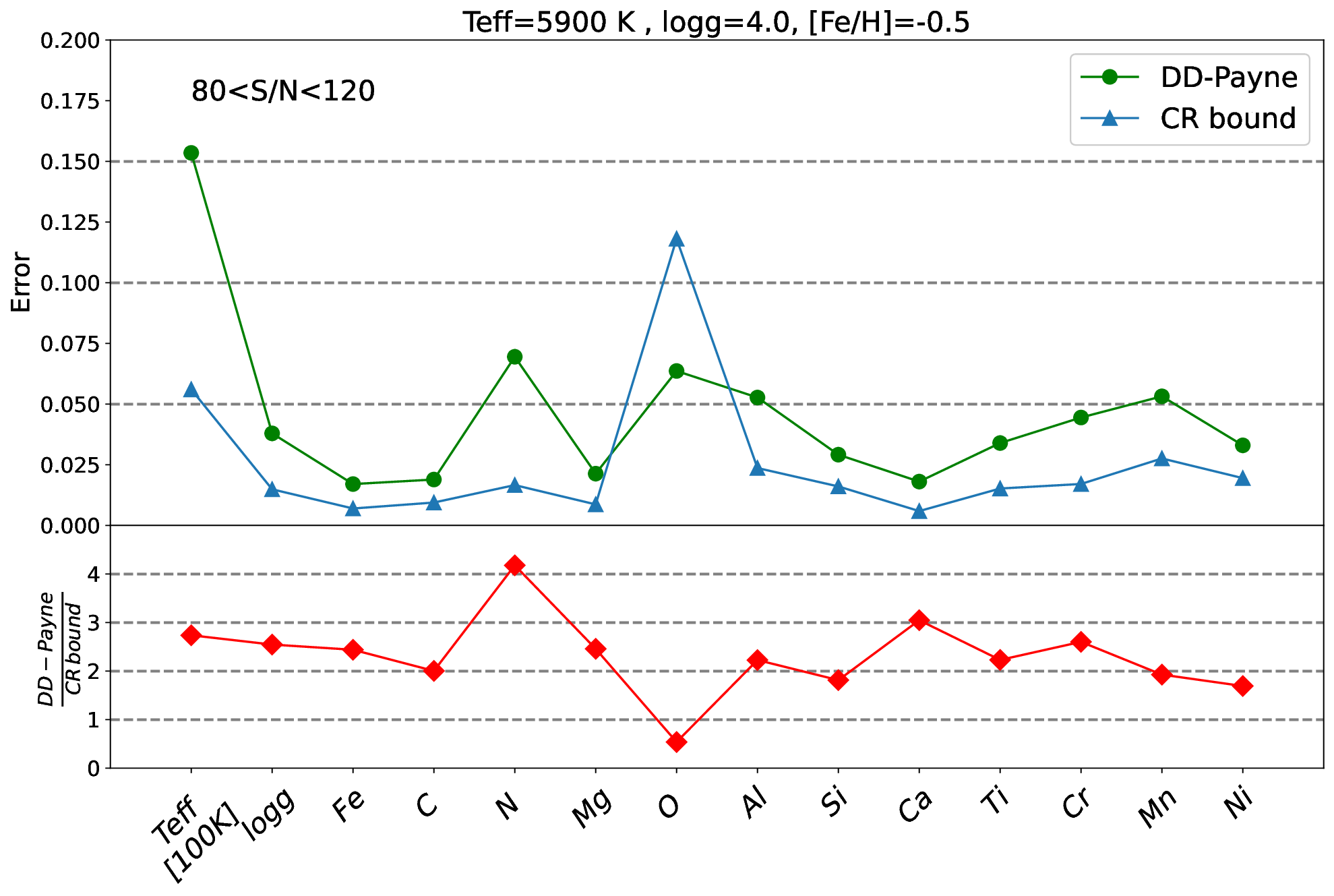}
\vspace{0.5em}
\caption{Comparison of the internal precision of the {\sc DD-Payne} stellar label estimates with theoretical precision limit calculated using the CR at a spectral $S/N$ of 100, for a fiducial star with $T_{\rm eff}=5900$~K, $\log~g=4.0$, and [Fe/H]=$-0.5$. For $T_{\rm eff}$, the value in the vertical axis has an unit of 100~K. The bottom panel shows the ratio between the {\sc DD-Payne} results and the CR bounds.
\label{fig8}}
\end{figure*}

Fig.~\ref{fig8} shows the measurement errors of the labels for stars with a fixed temperature $\teff\sim5900$~K, $\logg\sim4.0$, $\feh\sim-0.5$, and a spectral $S/N\sim100$. For comparison, the Cr\'amer-Rao \citep[CR;][]{Rao1945, Cramer1946} bound, which is a theoretical precision limit estimate of the DESI spectra, is also presented. For most of the labels, the measurement errors are around twice of the CR bound, except for the N and O. This is consistent with the case of LAMOST {\sc DD-Payne} results \citep{Sandford2020, Sandford2023}. The oxygen shows slightly smaller measurement error than the CR bound. The detailed reason is unclear, but we suspect it might be a consequence of covariance between other labels, especially between nitrogen and carbon \citep[e.g.][]{XiangMS2019}. Note that the above conclusion, i.e.,the measurement error is consistent with the CR bound within a factor of two, is also correct for lower S/N values as error is inverse function of S/N \citep{TingYS2017a}.

\subsection{Extinction, distance, and orbital parameters} 
\begin{figure*}[htb!]
\centering
\includegraphics[width=0.97\textwidth]{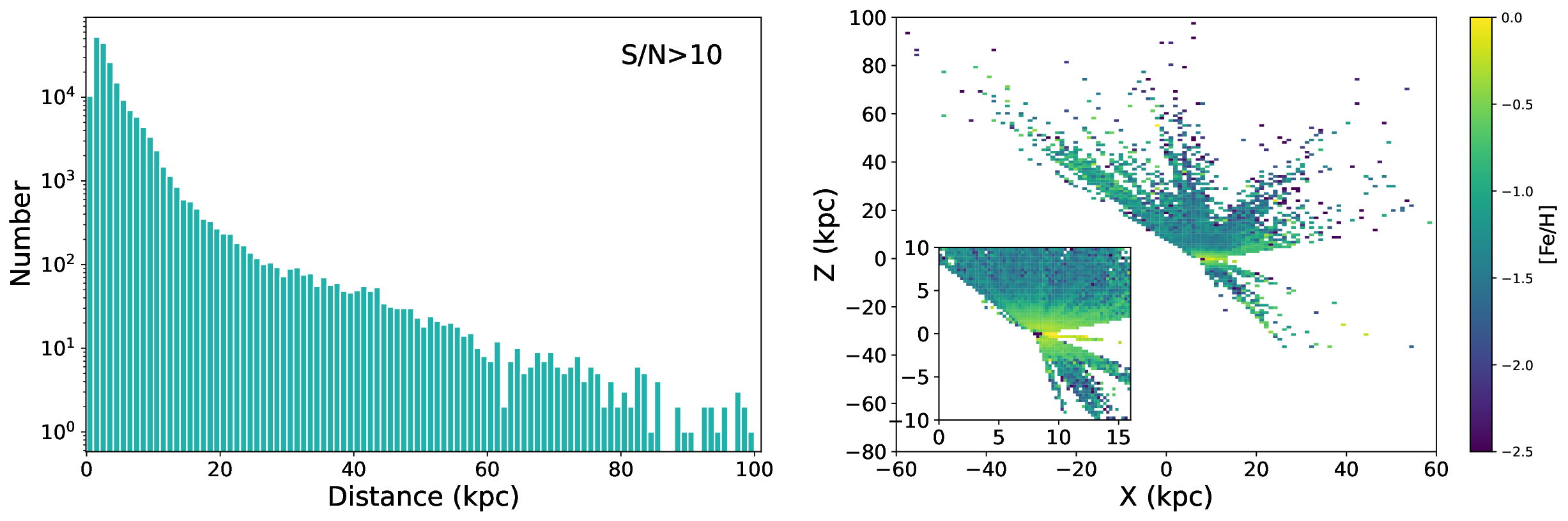}
\caption{{$\it Left$}: Distance distribution of the DESI EDR stars with spectral $S/N>10$. {$\it Right$}: Stellar distribution in the $X-Z$ plane of the Galactic Cartesian coordinate, color-coded by the median value of [Fe/H] in each $X-Z$ bin. The zoom-in plot shows the distribution in the window of $0<X<16$\,kpc and $-10<Z<10$\,kpc. 
\label{fig9}}
\end{figure*}
For convenience, we have also derived the extinction, distance, and orbital parameters of the sample stars. Combining the DESI stellar atmospheric parameters, the Gaia parallax and G/BP/RP photometry \citep{Gaia2023}, and the 2MASS J/H/K photometry \citep{Skrutskie2006}, we derive the extinction and distance with an empirical approach that similar to our previous work \citep[e.g.][]{YuanHB2015, XiangMS2022}. Here we only have a brief introduction of the methods. 

The extinction is derived with an empirical star-pair method \citep{YuanHB2013}. Firstly, we determine the intrinsic colors of a star based on its atmospheric parameters. The color excesses are then derived and converted to a reddening parameter $E_{B-V}$. To determine the intrinsic colors, we first define a set of control sample, which contains stars at high Galactic latitudes ($|b|>30^\circ$) that having small extinction values from the \citet[][hereafter SFD98]{Schlegel1998} extinction map. The intrinsic colors of stars in this control sample are derived by correcting for the SFD98 extinction. The intrinsic colors of the target stars are then derived by interpolating the control sample stars that have similar stellar parameters. To convert the color excesses for individual bands to $E_{B-V}$, and to derive the absolute extinction from $E_{B-V}$, we need an extinction coefficient for each single filter passband. We derived a $T_{\rm eff}-$ and [Fe/H]-dependent extinction coefficient for the Gaia G/BP/RP \citep{Riello2021} and 2MASS J/H/K \citep{Skrutskie2006} using the Kurucz synthetic spectra grid \citep{Castelli2003}, assuming the \citet{Fitzpatrick1999} extinction law with $R_V=3.1$. Typical uncertainty in the $E_{B-V}$ estimates is 0.02\,mag. Note that as most of the DESI targets are at high Galactic latitudes, the SFD98 map can be used as an alternative solution practically. However, our method is general and can be used to stars at low Galactic latitudes.  

With a similar empirical method, we also derive the absolute magnitudes in Gaia $G$ band, $M_G$, and 2MASS $K$ band, $M_K$, from the stellar atmospheric parameters. Here the control sample stars for absolute magnitude estimation are those having small parallax errors and presumed to be single objects according to their magnitude differences between spectroscopic and geometric estimates \citep[e.g.][]{Coronado2018, XiangMS2021}. Combining these absolute magnitude estimates with photometry and extinction, we derive a spectroscopic distance $\log{d_{\rm spec}}$ for each star using distance modulus,
\begin{align}
   m - M = 5\log{d_{\rm spec}} - 5 + A, 
\end{align}
where $A$ is the absolute extinction values in corresponding passbands. 
We use the $M_G$ if Gaia $G$ band photometry is available, otherwise the $M_K$ is adopted. This spectroscopic distance is combined with geometric distance $\log d_{\rm geom}$ derived by the inverse of the Gaia parallax, 
\begin{align}
   \log{d_{\rm geom}} = -\log\omega.
\end{align}
Here the Gaia parallax $\omega$ refers to that after correcting for the parallax zero point following \citet{Lindegren2021}. 
We combine $\log{d_{\rm spec}}$ and $\log{d_{\rm geom}}$ to give the final distance estimates with a weighted mean scheme by taking their invariance as the weight. In doing so, we apply a detailed treatment of possible high-mass ratio binaries whose $\log{d_{\rm spec}}$ could be erroneous due to light contribution from the secondary, similar to \citet{XiangMS2021}. For distant objects whose parallax-over-error is smaller than 2, or objects with a Gaia RUWE \citep{Lindegren2021b} larger than 1.4, we simply adopt the $\log{d_{\rm spec}}$ as the final distance estimates. 

The uncertainty in the relative distance estimates, i.e., $\sigma_d/d = {\rm ln}(10)\times\sigma_{\log{d}}$, decreases from 14\% for stars with $5<S/N<10$ to below 4\% for stars $S/N>50$. The overall sample have a median distance error of 9\%. The left panel of Fig.~\ref{fig9} shows the distance distribution for stars with a distance error smaller than 30\% and $S/N>10$, which allows a robust [Fe/H] determination. The sample stars spread a distance out to $\sim100$~kpc. Among them, 11,216 stars ($5.9\%$) are beyond 10\,kpc, and 1043 stars are beyond 35\,kpc.

Finally, we have also derived the orbits of sample stars from the distance derived above, the Gaia DR3 proper motion, and the DESI EDR line-of-sight (radial) velocity estimates, utilizing the $Galpy$ module (\citealt{Bovy2015}) and assuming the `MWPotential2014' Milky Way potential. The right panel of Fig.~\ref{fig9} shows the distribution of sample stars in the $X-Z$ plane of the Galactic Cartesian coordinate. The figure shows a clear negative metallicity trend in the vertical direction of the Galactic disk plane, as expected.  

\subsection{The Gaia-Enceladus-Sausage} 
\begin{figure*}[htb!]
\centering   
\subfigure
{
	\begin{minipage}{0.48\linewidth}
	\centering  
	\includegraphics[width=0.97\columnwidth]{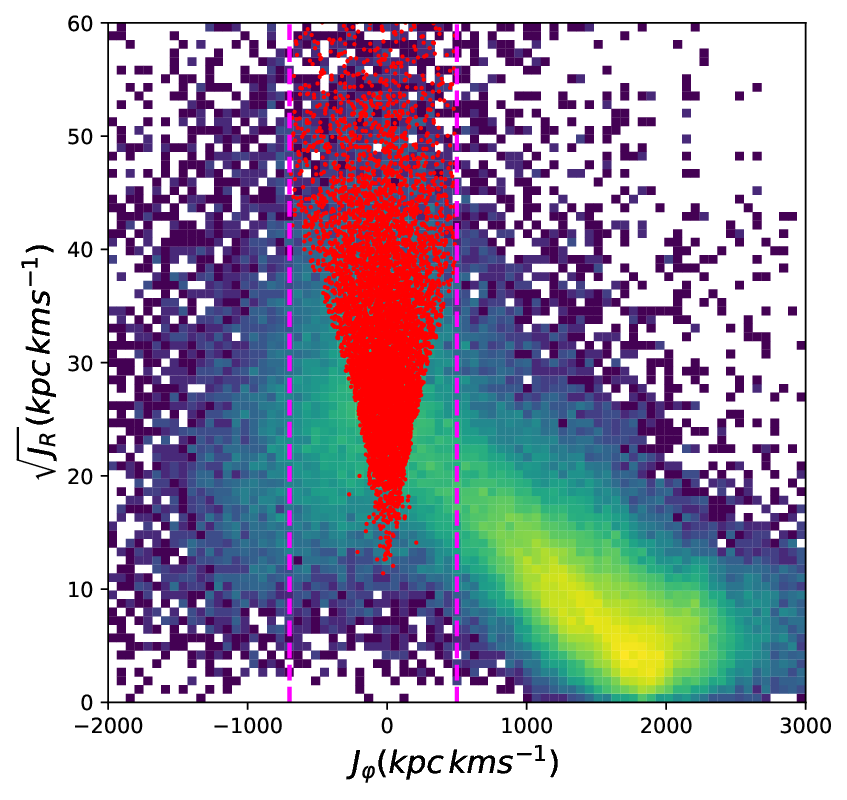} 
	\end{minipage}
} 
\subfigure
{
	\begin{minipage}{0.48\linewidth}
	\centering     
	\includegraphics[width=0.97\columnwidth]{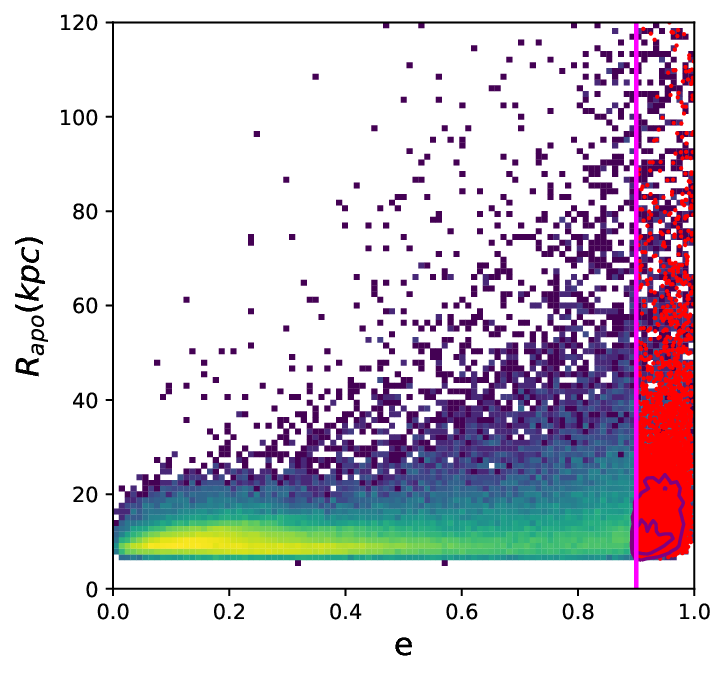}  
	\end{minipage}
}
\subfigure
{
	\begin{minipage}{0.48\linewidth}
	\centering  
	\includegraphics[width=0.97\columnwidth]{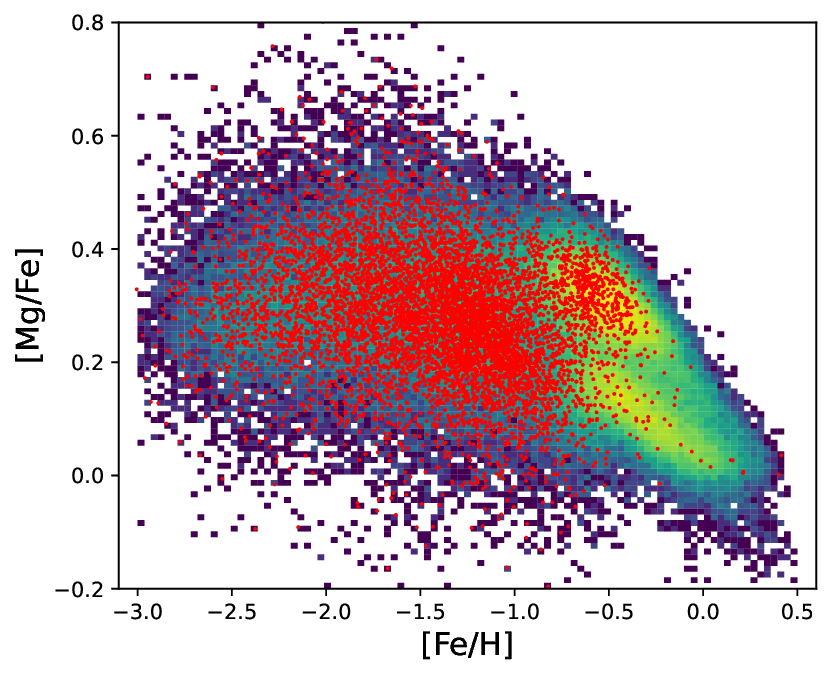} 
	\end{minipage}
} 
\subfigure
{
	\begin{minipage}{0.48\linewidth}
	\centering     
	\includegraphics[width=0.97\columnwidth]{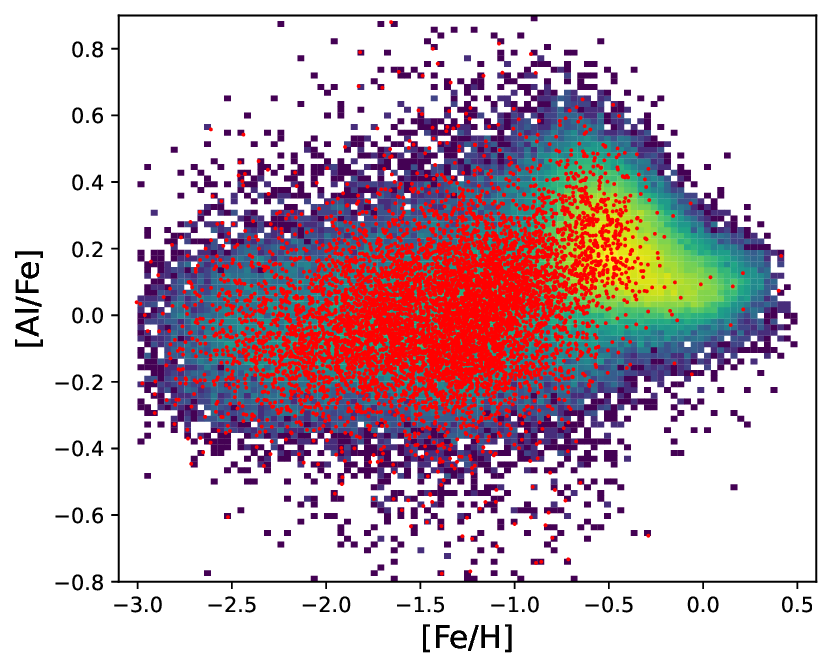}  
	\end{minipage}
}
\caption{Stellar number density distributions in the $J_\phi$-$J_R$  action plane (top-left) and the apo-center radius ($R_{\rm apo}$) - orbital eccentricity ($e$) plane (top-right). Colors represent the number density in individual bins of the plot. Vertical lines in both plots delineate the criteria we adopt to select the Gaia-Enceladus-Sausage (GES) stars via kinematic criteria (see text for details). The resultant low-angular momentum and high-eccentricity GES stars are shown with red dots. The black lines in the top-right panel show two equal-density contours of the red dots. The bottom panels show the stellar number density distribution in [Mg/Fe]-[Fe/H] (bottom-left) and the [Al/Fe]-[Fe/H] (bottom-right) planes, with (kinematic) GES stars marked in red dots (see text for details). 
\label{fig10}}
\end{figure*}

The vast majority of stars in the Galactic inner halo were revealed to be remnants of a satellite galaxy, dubbed as Gaia-Sausage \citep{Belokurov2018} or Gaia-Enceladus \citep{Helmi2018} that merged with our Milky Way in ancient time. For convenience, here we combine these nomenclatures to be the Gaia-Encelads-Sausage name (GES), similar to previous work \citep[e.g][]{Carrillo2022}. GES stars have low orbital angular momentum ($L_Z\simeq0$, or $e\simeq1.0$), and low [$\alpha$/Fe] values \citep{Belokurov2018, Helmi2018, Helmi2020}. A variety of criteria have been proposed to select GES stars from either kinematics or chemistry \citep{Haywood2018, DiMatteo2019, Feuillet2020, Helmi2020, Myeong2018,Naidu2020, Bonaca2020,Conroy2022}. Usually, different criteria lead to different fraction of the halo stars that belong to GES, ranging from 20\% to 80\% \citep[e.g.,][]{ Deason2018, Necib2019, WuWB2022}. Regardless the detailed number of fractions, it is no doubt that GES is a most prominent structure in the chemistry and kinematic spaces for the inner halo stellar populations.  

The upper panels of Fig.~\ref{fig10} show the distribution of the DESI EDR stars in the $J_\phi-J_R$ and $e-R_{\rm apo}$ planes, where $J_\phi$ and $J_R$ are the azimuthal and radial actions, respectively. $e$ is the orbital eccentricity, and $R_{\rm apo}$ the orbital apocenter radius. We select GES stars in this DESI EDR sample via the kinematic criteria 
\begin{equation}
  \begin{cases}
    -700 < J_\phi < 500\, {\rm km/s~kpc}, {\rm and} \\
    e>0.9, \\
    \end{cases}
\end{equation}
These criteria leads to 7148 stars, which are shown as red dot in Fig.~\ref{fig10}.  
The selected GES stars are shown in the chemical plane in bottom panels of Fig.~\ref{fig11}. The figure shows that the vast majority of the kinematically selected GES stars are distributed along a metal-poor, low-[$\alpha$/Fe] sequence, as expected (see Fig.~\ref{fig6}). Their [Al/Fe] values are also systematically lower than the relatively metal-rich, disk stars, which is consistent with previous work \citep{Das2020, Belokurov2022,Rix2022}. In addition, there is a bump of metal-rich ($\feh\simeq-0.5$), high-[$\alpha$/Fe] that share the same chemistry as the disk population. These are the well-known so-called `splashed' population, which was originally formed as disk population but turns into (kinematical) halo population due to ancient merger events \citep{Bonaca2020, Belokurov2022}.
These results verify the ability of using DESI stellar abundances to distinct accreted and in-situ stellar populations. 

\section{Discussion}
\subsection{Synergy with other spectroscopic surveys}
\begin{figure*}[htb!]
\centering   
\includegraphics[width=0.97\textwidth]{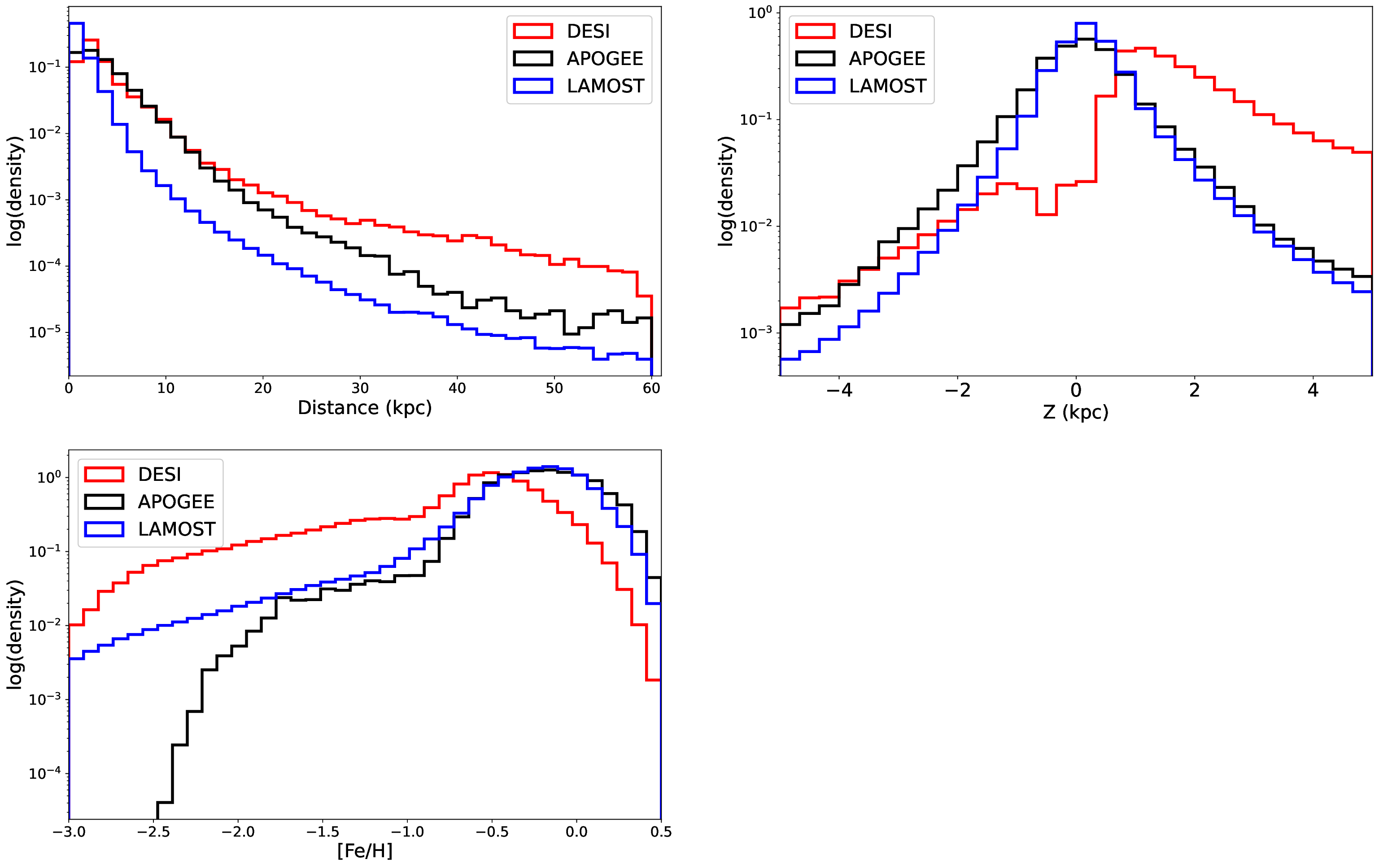}
\caption{Comparison of distance (top-left), height (top-right), and metallicity (bottom) distributions for DESI EDR, LAMOST DR9, as well as APOGEE DR17 stellar samples. The y-axis shows the normalized number density in logarithmic scale for both panels. For the DESI EDR sample, stars with $S/N>10$ are shown, while for LAMOST DR9 and APOGEE DR17, stars with $S/N>15$ are shown. The [Fe/H] values of LAMOST DR9 are determined with the {\sc DD-Payne} method (Zhang et al. in prep.), similar to this work. The DESI EDR sample contains a larger proportion of distant ($>20$~kpc), and metal-poor ($\feh<-1.0$) stars than the LAMOST and APOGEE samples, while the latter have better sampling for the nearby Galactic disk.
\label{fig11}}
\end{figure*}
Compared to other large surveys such as LAMOST and APOGEE, the DESI EDR samples fainter stars ($16<r<20$), thus reach further distance to the halo \citep{Cooper2023}. With a similar method to above, we have also estimated the distance of the LAMOST DR9 sample stars and the APOGEE DR17 sample stars. The upper-left panel of Fig.\,\ref{fig11} compares the distance distribution among the DESI, APOGEE, and LAMOST. It shows that DESI has a significantly higher fraction of stars beyond 20~kpc than the other two surveys, making it a valuable data set for studying the outer halo, especially after a larger data set is collected as the survey goes on. On the other hand, the upper-right panel of Fig.\,\ref{fig11} shows that the DESI sample contains only a tiny portion of nearby disk stars ($Z\lesssim2$~kpc), which is a manifest of the survey's sky footprint selection in Galactic latitudes \citep[e.g.][see also Fig.~\ref{fig9}]{Cooper2023}.   

The bottom panel of Fig.\,\ref{fig11} compares the metallicity distribution among those surveys. About 30\% of the DESI EDR stars are metal-poor ones ($\feh<-1.0$), while this proportion for LAMOST DR9 stars is 5 percent, and for the APOGEE DR17 is 3 percent. This result is consistent with \citet{Cooper2023}.
Note that the $\feh$ values for the LAMOST DR9 sample shown in the figure are determined by the {\sc DD-Payne} method (Zhang et al. in prep.). The figure shows a considerable fraction of very metal-poor stars with $\feh<-2.5$ also in the LAMOST sample. This is different from Figure~21 of \citet{Cooper2023}, who adopted the LAMOST DR7 official [Fe/H] release and observed a cut off at $\feh\simeq-2.5$.

Large surveys implemented on space telescopes such as CSST \citep{ZhanH2018, ZhanH2021} and Euclid \citep{Euclid2022} deliver multi-band photometry and slitless spectra covering a wavelength range from ultraviolet to infrared for faint stars down to $r\gtrsim25$~mag. Particularly, the CSST plans to implement an ultraviolet and optical ($255-1000$~nm) survey over a sky area of more than $15,000$ square degrees for $|b|>20^\circ$. The survey will take multi-band photometry for objects with $17\lesssim r\lesssim26$~mag, and slitless spectra ($R\sim200$) for objects with $15\lesssim r\lesssim22$~mag. For stellar label estimation of these deep photometric and low-resolution spectroscopic surveys, DESI stellar sample can serve as a golden source of training sets due to its superior magnitude depth and high sampling density in a large sky area towards the Galactic halo. 
 
\subsection{Future improvement of the method}
A robust performance of the {\sc DD-Payne} label estimation relies on a good training set. In the current work, we have adopted stellar labels from the LAMOST spectra, which have a lower spectral resolution power than DESI spectra, to be the training labels as a compromise due to the lack of sufficient training labels from high-resolution spectra. This situation can be improved in future as the survey collects more spectra that in common with high-resolution surveys, such as SDSS-V \citep{Kollmeier2017}.  

The DESI spectra have the potential to derive robust abundances estimates down to $\feh\lesssim-4$ for a few key elements such as Fe, C, and Ca. Our current gradient spectra used for regularization of the {\sc DD-Payne} only reach $\feh\simeq-3$ in the metal-poor side. This is expected to be further extended to more metal-poor case in our future work to maximally exploit the potential of DESI spectra. 

The DESI EDR sample contains mostly main sequence and main-sequence turn-off stars, whereas there are only a small number of red giant stars that can reach a far distance.
This is expected to be improved significantly in future DESI data releases \citep{Cooper2023}. 
\section{Conclusion}
Stellar abundances are crucial for unraveling the Galactic formation and evolution history. We have applied the {\sc DD-Payne}, a data-driven method regularised by differential spectra from stellar physical models, to the DESI EDR spectra for the determination of stellar atmospheric parameters and individual elemental abundances. In doing so, DESI EDR stars in common with APOGEE DR17 and LAMOST DR9 are adopted as the training set. Synthetic differential spectra computed with the ATLAS12 model atmosphere are adopted for regularizing the training process of {\sc DD-Payne}.

Our implementation derives $T_{\rm eff}$, $\log~g$, $v_{\rm mic}$, $\feh$ and abundance ratios [X/Fe] for X = {C, N, O, Mg, Al, Si, Ca, Ti, Cr, Mn, Ni} for 520,228 stars. The $T_{\rm eff}$, $\log~g$, and $\feh$ estimates are in good agreement with those in the DESI EDR catalog, except for the fact that a minor portion of stars in the latter suffer artificial grid effects. For spectra with $S/N>20$, the differences exhibit an overall dispersion of $\sim60$\,K in $T_{\rm eff}$, $\sim0.17$\,dex in $\log~g$, and 0.06\,dex in $\feh$, and a systematic deviation of comparable amounts.  

Our results demonstrated that rigorously determining abundances for more than 10 individual elements from the DESI spectra is feasible. For spectra with $S/N=100$, formal errors in the abundance determinations are smaller than 0.05 for most of the elements, and even smaller than 0.03 for a few elements, such as C, Mg, Ca, and Fe. These values are consistent with theoretical precision limits from the Cr\'amer-Rao bound calculation within a factor of two. 
The sample stars exhibit similar trends in most of the abundance spaces to those of the APOGEE high-resolution results. The high-$\alpha$ and low-$\alpha$ disk populations are well separated in the [Mg/Fe]-[Fe/H] plane, similar to previous studies. The Gaia-Enceladus-Sausage stars, defined by their orbital properties, are also discernible from the in-situ halo and disk populations in the [Mg/Fe]-[Fe/H] and [Al/Fe]-[Fe/H] planes. These results demonstrates the capability of disentangling different stellar populations of Galactic disk and halo with our abundance estimates.

We also derived the extinction, distance, and orbital parameters for the sample stars. The stars spread a distance out to $\sim$100~kpc, with a significant higher fraction of distant (or metal-poor) stars compared to other existed spectroscopic surveys, making it a power data set to study the Galactic outskirts. All these parameters are made public available, and serve as a useful complement to the DESI EDR catalog. Given its superior in survey depth, the DESI stellar sample can also serve as a golden source of training set for stellar label estimation of deep multi-band photometric and slitless spectroscopic surveys such as CSST and Euclid. 
 
As the survey goes on, we expect to have better training sets as there will be more common stars between DESI and other surveys, e.g., the APOGEE. So that the {\sc DD-Payne} method can be further improved, and we may deliver both better (higher precision, wider parameter space coverage) and larger data sample with future data releases.   
\newpage
\noindent {\bf Acknowledgments}
We acknowledges financial support from the National Key R\&D Program of China Grant (No.2022YFF0504200), and the National Natural Science Foundation of China (NSFC; Grant No.2022000083, No.12303025). 
Y.S.T. acknowledges financial support from the Australian Research Council through DECRA Fellowship DE220101520. 
H. Z. acknowledges the supports from the Beijing Municipal Natural Science Foundation (grant No. 1222028) and the NSFC (grant No. 12120101003, 12373010).

The DESI Legacy Imaging Surveys consist of three individual and complementary projects: the Dark Energy Camera Legacy Survey (DECaLS), the Beijing-Arizona Sky Survey (BASS), and the Mayall z-band Legacy Survey (MzLS). DECaLS, BASS and MzLS together include data obtained, respectively, at the Blanco telescope, Cerro Tololo Inter-American Observatory, NSF’s NOIRLab; the Bok telescope, Steward Observatory, University of Arizona; and the Mayall telescope, Kitt Peak National Observatory, NOIRLab. NOIRLab is operated by the Association of Universities for Research in Astronomy (AURA) under a cooperative agreement with the National Science Foundation. Pipeline processing and analyses of the data were supported by NOIRLab and the Lawrence Berkeley National Laboratory. Legacy Surveys also uses data products from the Near-Earth Object Wide-field Infrared Survey Explorer (NEOWISE), a project of the Jet Propulsion Laboratory/California Institute of Technology, funded by the National Aeronautics and Space Administration. Legacy Surveys was supported by: the Director, Office of Science, Office of High Energy Physics of the U.S. Department of Energy; the National Energy Research Scientific Computing Center, a DOE Office of Science User Facility; the U.S. National Science Foundation, Division of Astronomical Sciences; the National Astronomical Observatories of China, the Chinese Academy of Sciences and the Chinese National Natural Science Foundation. LBNL is managed by the Regents of the University of California under contract to the U.S. Department of Energy. The complete acknowledgments can be found at \url{https://www.legacysurvey.org/}.

This work has made use of data products from the Guo
Shou Jing Telescope (the Large Sky Area Multi-Object Fibre Spectroscopic Telescope, LAMOST). LAMOST is a National Major Scientific Project built by the Chinese Academy of Sciences. Funding for the project has been provided by the National Development and Reform Commission. LAMOST is operated and managed by the National Astronomical Observatories, Chinese Academy of Sciences.

This work also has made use of data from the European Space Agency (ESA) mission Gaia \footnote{https://www.cosmos.esa.int/gaia}, processed by the Gaia Data Processing and Analysis Consortium (DPAC\footnote{https://www.cosmos.esa.int/ web/gaia/dpac/consortium}). Funding for the DPAC has been provided by national institutions, in particular the institutions participating in the Gaia Multilateral Agreement.

The following software and programming languages made this research possible: TOPCAT \citep{Taylor2005}; PYTHON (version 3.7) and accompany packages ASTROPY \citep{Astropy2013}, SCIPY \citep{Virtanen2020}, MATPLOTLIB \citep{Hunter2007}, NUMPY \citep{vanderWalt2011}, and PayTorch \citep{Paszke2019}. 
\newpage
\bibliographystyle{aasjournal}
\bibliography{desi.bib}
\end{document}